\documentclass[aps,prl,twocolumn,superscriptaddress,nopacs,a4paper]{revtex4}

\usepackage{graphicx}

\pdfoutput=1 

\usepackage{fancyhdr}
\usepackage[colorlinks=true,citecolor=blue,linkcolor=magenta]{hyperref}
\hypersetup{%
pdfauthor = {Patrick Maletinsky},
colorlinks = true, linkcolor = blue, urlcolor=blue, bookmarksnumbered =  true}

\begin{document}


\title{A robust, scanning quantum system for nanoscale sensing and imaging}


\author{P. Maletinsky\footnote{These authors contributed equally to this work}}
\affiliation{Department of Physics, Harvard University, Cambridge, Massachusetts 02138 USA}
\email[]{patrickm@physics.harvard.edu}
\author{S. Hong\footnotemark[\value{footnote}]}
\affiliation{School of Engineering and Applied Science, Harvard University, Cambridge, Massachusetts, 02138 USA}
\author{M.S. Grinolds\footnotemark[\value{footnote}]}
\affiliation{Department of Physics, Harvard University, Cambridge, Massachusetts 02138 USA}
\author{B. Hausmann}
\affiliation{School of Engineering and Applied Science, Harvard University, Cambridge, Massachusetts, 02138 USA}
\author{M.D.Lukin}
\affiliation{Department of Physics, Harvard University, Cambridge, Massachusetts 02138 USA}
\author{R.-L. Walsworth}
\affiliation{Department of Physics, Harvard University, Cambridge, Massachusetts 02138 USA}
\affiliation{Harvard-Smithsonian Center for Astrophysics, Cambridge, Massachusetts 02138 USA}
\author{M. Loncar}
\affiliation{School of Engineering and Applied Science, Harvard University, Cambridge, Massachusetts, 02138 USA}
\author{A. Yacoby}
\affiliation{Department of Physics, Harvard University, Cambridge, Massachusetts 02138 USA}
\email[]{yacoby@physics.harvard.edu}

\date{\today}

\begin{abstract}
Controllable atomic-scale quantum systems hold great potential as sensitive tools for nanoscale imaging and metrology\,\cite{Sekatskii1996,Michaelis2000,Chernobrod2005,Balasubramanian2008,Maze2008a,Taylor2008}.
Possible applications range from nanoscale electric\,\cite{Dolde2011} and magnetic field sensing\,\cite{Balasubramanian2008,Maze2008a,Taylor2008,Degen2008} to single photon microscopy\,\cite{Sekatskii1996,Michaelis2000}, quantum information processing\,\cite{Neumann2010},
and bioimaging\,\cite{McGuinness2011}. At the heart of such schemes is the ability to scan and accurately position a robust sensor within a few nanometers of a sample of interest, while preserving the sensor's quantum coherence and readout fidelity. These combined requirements remain a challenge for all existing approaches that rely on direct grafting of individual solid state quantum systems\,\cite{Kuhn2001,Balasubramanian2008,Cuche2010} or single molecules\,\cite{Michaelis2000} onto scanning-probe tips. Here, we demonstrate the fabrication and room temperature operation of a robust and isolated atomic-scale quantum sensor for scanning probe microscopy. Specifically, we employ a high-purity, single-crystalline diamond nanopillar probe containing a single Nitrogen-Vacancy (NV) color center. We illustrate the versatility and performance of our scanning NV sensor by conducting quantitative nanoscale magnetic field imaging and near-field single-photon fluorescence quenching microscopy. In both cases, we obtain imaging resolution in the range of $20~$nm and sensitivity unprecedented in scanning quantum probe microscopy.
\end{abstract}

\maketitle

\begin{figure}
\includegraphics{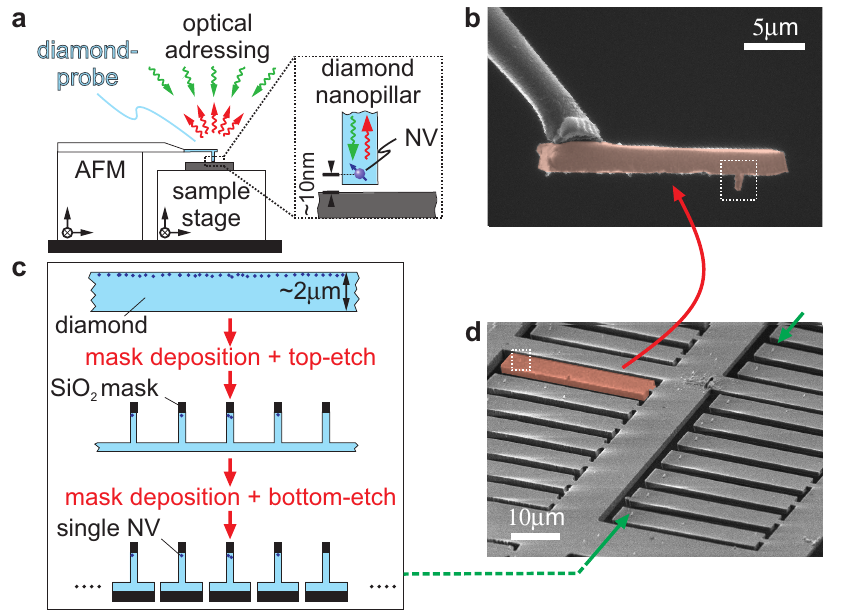}
\caption{\label{figureSchematics} \textbf{Experimental setup and probe fabrication for the scanning NV sensor}. (a) Schematic of the setup consisting of a combined optical and atomic force microscope (AFM). We use a $532~$nm laser (green arrows) to address the scanning NV center through its red fluorescence (red arrows). The scanning NV center resides in a diamond nanopillar (inset) and its proximity to the sample is maintained through AFM feedback. (b) Scanning electron microscope (SEM) image of a single-crystalline diamond nanopillar-probe (false-color coded in red) with a single NV center in its tip (see Fig.\,\ref{figureBasicProperties}). (c) Brief depiction of the fabrication process for scanning single-crystalline diamond NV sensors. Electron-beam lithography is used to define nanopillars and platforms from the top- and bottom-sides of a few micron thin diamond membrane. Patterns are then transferred to the diamond by reactive ion etching. (d) SEM image of a finalized array of diamond platforms with nanopillars. In all panels, dotted rectangles highlight diamond nanopillars.}
\end{figure}

The NV center in diamond is a point-defect that offers the potential for sensing and imaging with atomic scale resolution. Sensitive nanoscale detection of various physical quantities is possible because the NV center forms a bright and stable single photon source\,\cite{Kurtsiefer2000} for optical imaging, and possesses a spin-triplet ground state which offers excellent magnetic\,\cite{Maze2008a} and electric\,\cite{Dolde2011} field sensing capabilities. The remarkable performance of the NV center in such spin-based sensing schemes, is the result of the long NV spin coherence time\,\cite{Balasubramanian2009}, combined with efficient optical spin preparation and readout\,\cite{Gruber1997}, all at room temperature. In addition, NV centers can be positioned within nanometers of a diamond surface\,\cite{Bradac2010} and therefore in close proximity of a sample to maximize signal strengths and spatial resolution. In order to realize the full potential of these attractive features, we have developed a "scanning NV sensor" (Fig.\,\ref{figureSchematics}a), which employs a diamond nanopillar as the scanning probe, with an individual NV center artificially created within a few nanometers of the pillar tip through ion implantation. Long NV spin coherence times ($\approx30~\mu$s) are achieved as our devices are fabricated from high purity, single-crystalline bulk diamond\,\cite{Stanwix2010}. Furthermore, diamond nanopillars are efficient waveguides for the NV fluorescence band\,\cite{Babinec2010}, which yields record-high NV signal collection efficiencies for a scanning NV device.

Fig.\,\ref{figureSchematics}b shows a representative scanning electron microscope (SEM) image of a single-crystalline diamond scanning probe containing a single NV center. The preparation of such devices is based on recently developed techniques in diamond nanofabrication\,\cite{Hausmann2010}, combined with established methods for controlled NV creation through ion implantation\,\cite{Kalish1997}. Our scanning diamond nanopillars have typical diameter $\approx200~$nm and length of $1~\mu$m and are fabricated on few-micron sized diamond platforms which can be attached to atomic force microscope (AFM) tips for scanning (Fig.\,\ref{figureSchematics}b and  methods). Our fabrication procedure (Fig.\,\ref{figureSchematics}c) allows for highly parallel processing as shown in the array of diamond devices depicted in the SEM image in Fig.\,\ref{figureSchematics}d. From this array, we select nanopillars that contain single NV centers with high photon count rates and long spin coherence times and mount these single-NV nanopillars onto AFM tips to yield the finalized scanning probe shown in the SEM picture in Fig.\,\ref{figureSchematics}b.

To employ the scanning NV sensor and characterize its basic spin and optical properties, we used a combined confocal- and atomic force-microscope as sketched in Fig.\,\ref{figureSchematics}a. The setup is equipped with piezo positioners for the sample and AFM-probe to allow for independent scanning with respect to the optical axis. Optical addressing and readout of the NV center in the tip is performed through a long working-distance microscope objective (numerical aperture, NA$=0.7$), to accommodate the AFM-head between the sample and objective. Here, an important advantage of employing diamond nanopillars for scanning is their property to collimate NV emission (resulting in a low exit-NA $\approx 0.65$\,\cite{Hausmann2010}), yielding a high collection efficiency even with low NA collection-optics.

\begin{figure}
\includegraphics{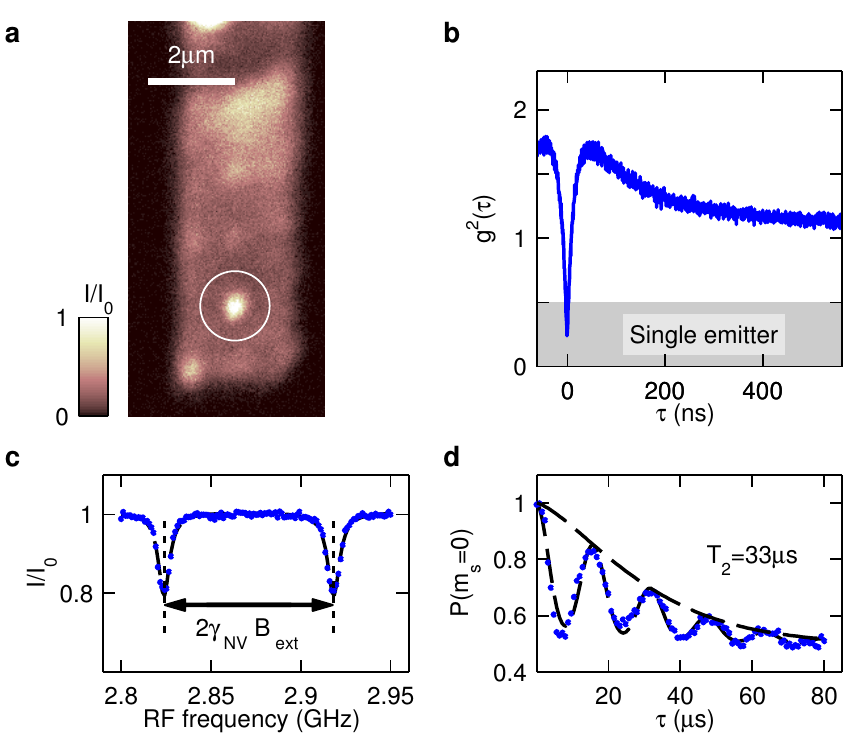}
\caption{\label{figureBasicProperties} \textbf{Single NV centers in scanning diamond nanopillars}. (a) Confocal image of red fluorescence from a single-crystalline diamond probe (see side view SEM image in\,\ref{figureSchematics}b). Fluorescence counts are normalized to $I_0=1.5\cdot10^5~$cps. The encircled bright feature stems from fluorescence of a single NV center in a nanopillar. (b) Photon autocorrelation measurement (g$^2(\tau)$) for NV fluorescence in a scanning nanopillar device. Data with g$^2<0.5$ (grey-shaded region) demonstrates the presence of a single photon emitter in the nanopillar. (c) Optically detected electron spin resonance (ESR) identifies the single emitter in the nanopillar as an NV center. The two possible NV spin transitions\,\cite{Gruber1997} are split by the NV electron Zeeman splitting $2 \gamma_{\rm NV} B_{\rm NV}$, where $\gamma_{\rm NV}=2.8~$MHz/G is the NV gyromagnetic ratio and $B_{\rm NV}$ is the magnetic field along the NV axis (here, $B_{\rm NV}\approx16~$G). (d) Spin-echo measurement for an NV center on a diamond nanopillar device. The envelope fitted to the characteristic NV spin-echo decay (see methods) yields the NV spin-coherence time of $T_2=33~\mu$s. Data in panels a-d was each taken on different devices with similar properties.}
\end{figure}

Fig.\,\ref{figureBasicProperties}a shows a confocal scan under green laser illumination ($\lambda_{\rm exc}=532~$nm) of a typical single NV/AFM device. The bright photon emission emerging from the nanopillar (white circle) originates from a single NV center, as evidenced by the pronounced dip in the photon-autocorrelation measurement (Fig.\,\ref{figureBasicProperties}b) and the characteristic signature of optically detected NV electron-spin resonance (ESR)\,\cite{Gruber1997} (Fig.\,\ref{figureBasicProperties}c), all obtained on different devices.
Importantly, we found that photon waveguiding through the nanopillar dramatically increases  excitation and detection efficiencies for NV fluorescence\,\cite{Babinec2010}. For some devices, maximal NV fluorescence count-rates exceeding $3\cdot10^5~$counts per second (cps) were observed for excitation powers as low as $20~\mu$W. We thus significantly increase fluorescence signal-strength from the single scanning NV and at the same time minimize exposure of samples to green excitation light, which is especially relevant for possible biological or low-temperature applications of the scanning NV sensor.

Using well-established techniques for coherent NV-spin-manipulation\,\cite{Jelezko2004}, we characterized the spin-coherence time, T$_2$, of a single NV center in a diamond nanopillar. Spin-coherence sets the NV sensitivity to magnetic fields and limits the number of coherent operations that can be performed on an NV spin; it is therefore an essential figure of merit for applications in magnetic field imaging\,\cite{Taylor2008} and quantum information processing\,\cite{Neumann2010}. Using a Hahn-echo pulse sequence, we measured the characteristic single NV coherence decay\,\cite{VanOort1990} shown in Fig.\,\ref{figureBasicProperties}d; and from the decay envelope we deduce a spin-coherence time of T$_2=33~\mu$s. We note that this T$_2$ time is consistent\,\cite{Taylor2008} with the density of implanted Nitrogen ions ($3\cdot10^{-11}~$cm$^{-2}$) and conclude that our device fabrication procedure fully preserves NV spin coherence. Combining measurements of the T$_2$-time with the fluorescence count-rate and spin-readout contrast of the single NV in Fig.\,\ref{figureBasicProperties}d, yields an AC magnetic field sensitivity\,\cite{Taylor2008} of $170~$nT/$\sqrt{\rm Hz}$.

\begin{figure}
\includegraphics{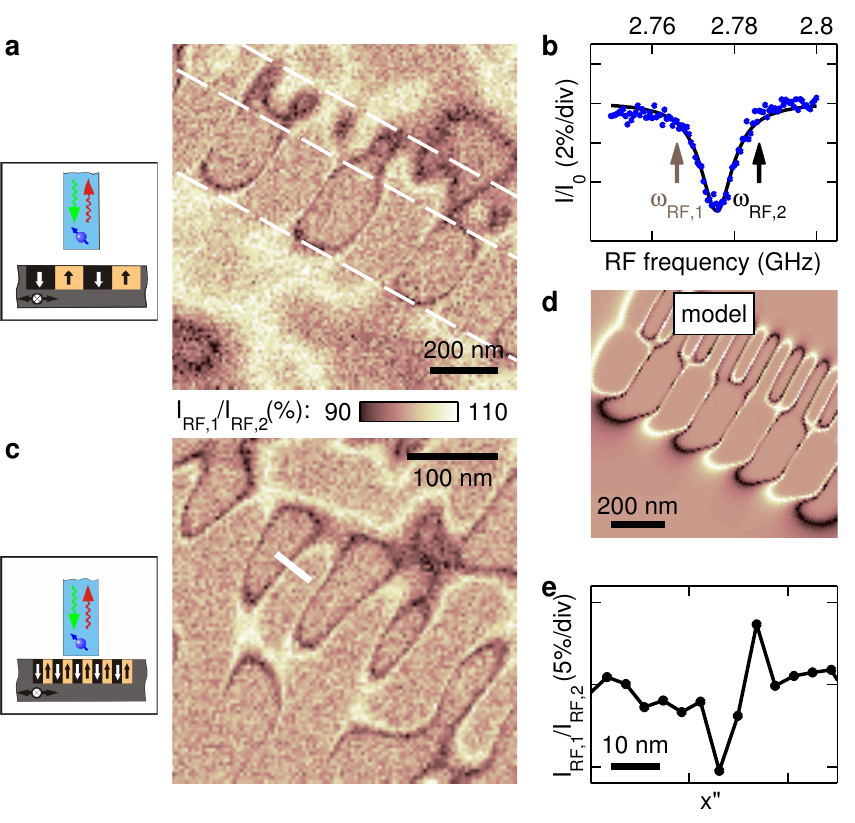}
\caption{\label{figureMAG} \textbf{Nanoscale magnetic field imaging with the scanning NV sensor}. (a) NV magnetic field image of tracks on a magnetic harddrive, highlighted by dashed white lines added to image. The image shows normalized data $I_{\rm RF,1}/I_{\rm RF,2}$ (see (b)) and thereby reveals magnetic field lines corresponding to $B_{\rm NV}=\pm 3~$G (dark and bright contours, respectively). Total image acquisition time was $11.2~$minutes. (b) Optically detected ESR resonance of the sensing NV center. For magnetic field imaging, we modulate an applied microwave field between two frequencies ($\omega_{\rm RF, 1}=2.766~$GHz and $\omega_{\rm RF, 2}=2.786~$GHz) and collect NV fluorescence counts ($I_{\rm RF,1}$ and $I_{\rm RF,2}$, respectively) in synchrony with the RF modulation. (c) Magnetic image obtained with the same method as in (a), but with a decreased NV-sample distance. Bringing the NV closer to the sample increases the magnetic field magnitude at the NV sensor, and improves the imaging spatial resolution, allowing imaging of $\approx30~$nm magnetic bits. (d) Calculated NV response for the experimental situation in (a), assuming a simplified magnetic sample (see Supplementary Information) (e) Linecut along the white line indicated in (c) (averaged across $6$ adjacent pixels). The scanning NV sensor's ability to spatially resolve magnetic fields is limited by the local magnetic field gradient, which for the present system realization and magnetic harddrive sample leads to a resolution $\approx 3~$nm. Insets in a and b illustrate the experimental configuration, with the sensing NV center fixed on the optical axis and the magnetic sample scanned below the pillar.}
\end{figure}

To demonstrate the resolving power of the scanning NV sensor in magnetic imaging\,\cite{Balasubramanian2008,Maze2008a}, we imaged a nanoscale magnetic memory medium consisting of bit-tracks of alternating (out-of-plane) magnetization with various bit-sizes. Fig.\,\ref{figureMAG} illustrates our method and results: the scanning NV sensor operated in a mode that imaged contours of constant magnetic field strength ($B_{\rm NV}$) along the NV axis through the continuous monitoring of red NV fluorescence, in the presence of an ESR driving field of fixed frequency $\omega_{\rm RF}$. We detuned $\omega_{\rm RF}$ by $\delta_{\rm RF}$ from the bare NV spin transition-frequency, $\omega_{\rm NV}$, but local magnetic fields due to the sample changed this detuning during image acquisition. In particular, when local fields brought the NV's spin-transition into resonance with $\omega_{\rm RF}$, we observed a drop in NV fluorescence rate, which in the image yielded a contour of constant $B_{\rm NV}=\delta_{\rm RF}/\gamma_{\rm NV}$, with $\gamma_{\rm NV}=2.8~$MHz/G being the NV gyromagnetic ratio. In order to reject low-frequency noise\,\cite{Grinolds2011} and image domains of opposite magnetization, we simultaneously acquired two such images by applying RF sidebands to $\omega_{\rm NV}$ with $\delta_{\rm RF}=\pm10~$MHz (dark and bright arrows in Fig.\,\ref{figureMAG}b). Normalization of the pixel-values in the two data-sets then directly provided a map of magnetic field contours with positive and negative values of $B_{\rm NV}$ (here, with $B_{\rm NV}=\pm3~$G). Fig.\,\ref{figureMAG}a shows such a scanning NV magnetometry image of two stripes of magnetic bits (indicated by the white dashed lines) with nominal bit-spacings of $125~$nm and $50~$nm. The shape of the observed domains is well reproduced by calculating the response of the NV magnetometer to an idealized sample with rectangular magnetic domains of dimensions corresponding to the written tracks (see Fig.\,\ref{figureMAG}d and Supplementary Information).

Spatial resolution in scanning-probe microscopy is limited by the distance of the probe to the sample.
Therefore, further approaching the NV sensor to the magnetic sample revealed magnetic bits with resolution of about $30~$nm, as shown in Fig.\,\ref{figureMAG}c. Even though in this image magnetic field-lines can be imaged with $\approx3~$nm resolution (Fig.\,\ref{figureMAG}e), smaller magnetic domains remained unresolved in Fig.\,\ref{figureMAG}c due to the remaining distance between the NV center in the nanopillar and the magnetic sample. For the particular sample under investigation, a further decrease of NV-to-sample distance was not beneficial as it severely reduced overall imaging contrast: Due to the sample's strong magnetization, significant local magnetic fields transverse to the NV axis led to a reduction of NV fluorescence intensity and ESR contrast. We note that while this effect suppressed the visibility of magnetic field lines, local modifications of fluorescence intensity could still be used to image magnetic bits (see Supplementary Information). In addition to these magnetic effects, the sample's metallic nature caused strong fluorescence quenching\,\cite{Buchler2005} when the NV center was brought close to the sample, which further reduced the signal-to-noise for magnetic sensing.

\begin{figure}
\includegraphics{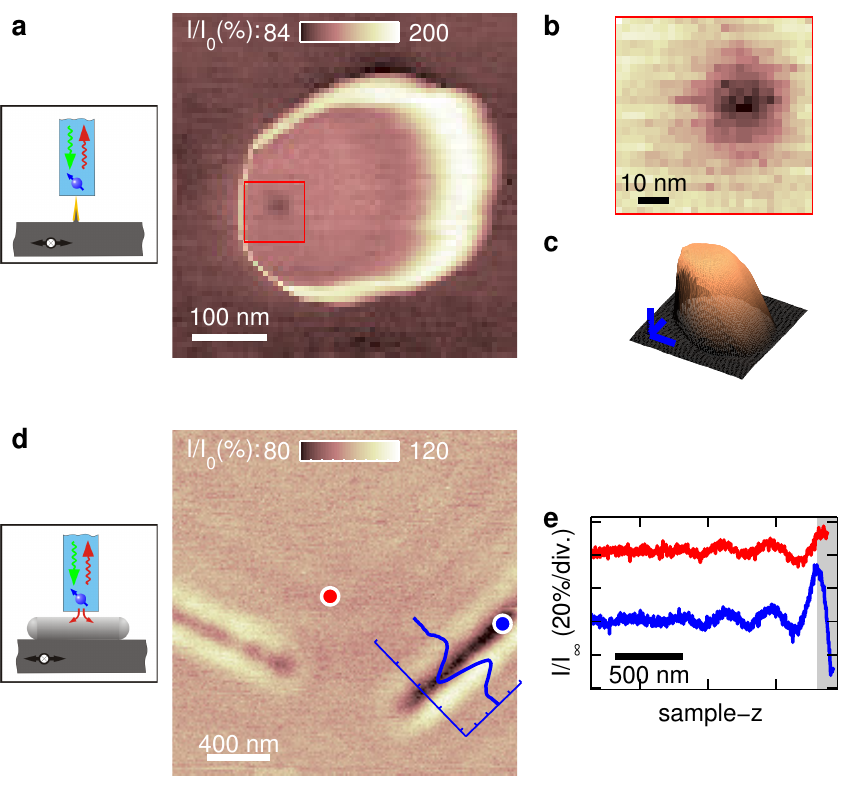}
\caption{\label{figureFluoQuench} \textbf{Nanoscale fluorescence quenching imaging (FQI) using the scanning NV sensor}. (a) FQI of a sharp metallic tip on a dielectric substrate. Scanning the diamond pillar over a sharp tip leads to a bright, circular feature due to sample-topography (see Supplementary Information). Positioning the metallic tip exactly at the location of the NV center (red square), however, yields a sharp dip in NV fluorescence. (b) Zoomed-in image of the red square region in (a); the observed fluorescence quenching dip demonstrates an FQI spatial resolution $\approx~20$nm. (c) AFM topography image obtained simultaneously with the data in (b). Blue scalebars represent $100~$nm displacement in all directions. (d) Scanning FQI of silver nanowires dispersed on a quartz substrate. The inset shows a linecut across a wire, indicating resolution of the intrinsic wire diameter $\approx~100~$nm (horizontal (vertical) tick-spacing is $200~$nm ($10~$\%)). (e) Comparison of the NV fluorescence rate as a function of NV-sample distance on (red) and off (blue) a nanowire (see red and blue dots in FQI image). The data demonstrates that imaging contrast in FQI is acquired in the optical near-field of the NV and therefore enables a breaking of the diffraction-limit. Red and blue curves are offset by $40~$\% for clarity. Insets in a and d illustrate the experimental configuration in FQI.}
\end{figure}

Sample-induced quenching of NV fluorescence, while adverse to magnetic field sensing, opens  additional avenues for nanoscale imaging with scanning NV centers. In particular, NV centers are single photon sources, and as such can be utilized for near-field optical imaging beyond the diffraction and shot-noise limits\,\cite{Michaelis2000}, as well as for scanning ``F\"orster resonance energy transfer'' microscopy\,\cite{Sekatskii1996}. We used our scanning NV sensor to demonstrate such near-field optical imaging by performing scanning fluorescence-quenching imaging (FQI) on nanoscale metallic objects. Imaging contrast consisted of the detected NV fluorescence in the far-field changing when the NV was in proximity to a metallic object\,\cite{Buchler2005}.

We determined the scanning-NV FQI spatial resolution by measuring the point spread function (PSF) of our single photon microscope. To that end, we fabricated metallic tips with $< 20~$nm diameter (see schematic in Fig.\,\ref{figureFluoQuench}a and Supplementary Information), which we imaged by monitoring the total NV fluorescence rate as we scanned the sample in direct contact with the pillar (Fig.\,\ref{figureFluoQuench}a). The resulting data show signatures of the topography of the scanning diamond nanopillar (bright ring in the NV fluorescence signal, see Supplementary Information for details). More importantly, however, while the sharp metallic tip scanned the front-end of the diamond probe, we observed a pronounced dip in NV fluorescence (red square in Fig,\,\ref{figureFluoQuench}a and zoomed image in Fig.\,\ref{figureFluoQuench}b) when the metallic tip was positioned at the location of the NV center. The Gaussian width (double standard deviation) of $25.8~$nm of this fluorescence quenching spot was likely still limited by the size of the metallic tip and therefore marks an upper bound to the imaging resolution in FQI. In addition to yielding the PSF of this imaging mode, such data provides the location of the single NV center within the nanopillar, which we can readily correlate with the simultaneously acquired topography of the device (Fig.\,\ref{figureFluoQuench}c).

As a second example application, Fig.\,\ref{figureFluoQuench}d shows an FQI image of two silver nanowires on quartz, where we easily resolve the intrinsic nanowire diameter $\approx~100~$nm. We note that such sub-diffraction optical imaging is feasible because the scanning NV center forms an atomic-scale light-source\,\cite{Kuhn2001}, whose optical near-field contains spatial frequencies exceeding the inverse optical wavelength. This near-field coupling can be observed by monitoring NV fluorescence intensity as a function of the distance between an FQI sample and the NV. Fig.\,\ref{figureFluoQuench}e shows a comparison of such fluorescence ``approach-curves'' on and off a nanowire, which demonstrates that the imaging signal is acquired within $\approx 100~$nm of the sample surface (grey shaded area).

For all imaging applications demonstrated in this paper, spatial resolution is limited by  NV-to-sample distance. The biggest uncertainty to this distance is vertical straggle in the NV implantation process, which is still poorly understood\,\cite{Grinolds2011}. Advances in the controlled creation of NV centers close to diamond surfaces should enable production of stable NV centers as close as $3~$nm from the nanopillar's tip\,\cite{Bradac2010}. Spatial resolution for scanning NV imaging could therefore be further improved by about one order of magnitude. We note that for magnetic field imaging, our current ability to resolve individual magnetic domains already equals the typical performance of alternative methods\,\cite{Wolny2010,Kohashi2009}, with the added advantages of being non-invasive and quantitative.

The magnetic field sensitivity we demonstrated with the scanning NV sensor compares well to the performance realized previously with single NV centers in ultrapure, bulk diamond samples\,\cite{Maze2008a}. Combined with the mechanical robustness and durability of our diamond probes (up to several weeks of scanning with the same tip), our results constitute a significant advance in scanning quantum probe microscopy and demonstrate the advantage of our method over alternative approaches\,\cite{Cuche2010,Balasubramanian2008}.

The scope of applications of the scanning quantum probe described here goes far beyond imaging and sensing. For example, our nanoscopic, scannable single photon source, could be used to controllably inject single plasmonic excitations into nanometallic structures\,\cite{Mollet2011} at well-defined locations, which would have broad impact to the field of nano-plasmonics. Additionally, our device forms an ideal platform to coherently couple the scanning NV spin to other spin systems such as P in Si\,\cite{Kane1998}, other NV centers, or carbon-based spin qubits\,\cite{Recher2010}, either by optical or magnetic coupling. Quantum information could thereby be transferred between a stationary qubit and our scanning NV center and from there to other qubit systems or single photons\,\cite{Togan2010}.


We thank B. D. Terris and N. Supper from Hitachi GST for providing the magnetic recording samples and T.G. Tiecke and J.D. Thompson for providing the silver nanowire sample imaged in our experiments.  P.M. acknowledges support from the Swiss National Science Foundation, S.H. thanks the Kwanjeong Scholarship Foundation for funding and M.S.G. is supported through fellowships from the Department of Defense (NDSEG program) and the NSF. This work was supported by NIST and DARPA and in part performed at the Center for Nanoscale Systems (CNS), a member of the National Nanotechnology Infrastructure Network (NNIN), which is supported by the National Science Foundation under NSF award no.$ECS-0335765$. CNS is part of Harvard University.

Correspondence and requests for materials should be addressed to A.Y. (email: yacoby@physics.harvard.edu)

\bibliographystyle{apsrev4-1}
\bibliography{Scanning_NV_arxiv}

\begin{thebibliography}{39}%
\makeatletter
\providecommand \@ifxundefined [1]{%
 \@ifx{#1\undefined}
}%
\providecommand \@ifnum [1]{%
 \ifnum #1\expandafter \@firstoftwo
 \else \expandafter \@secondoftwo
 \fi
}%
\providecommand \@ifx [1]{%
 \ifx #1\expandafter \@firstoftwo
 \else \expandafter \@secondoftwo
 \fi
}%
\providecommand \natexlab [1]{#1}%
\providecommand \enquote  [1]{``#1''}%
\providecommand \bibnamefont  [1]{#1}%
\providecommand \bibfnamefont [1]{#1}%
\providecommand \citenamefont [1]{#1}%
\providecommand \href@noop [0]{\@secondoftwo}%
\providecommand \href [0]{\begingroup \@sanitize@url \@href}%
\providecommand \@href[1]{\@@startlink{#1}\@@href}%
\providecommand \@@href[1]{\endgroup#1\@@endlink}%
\providecommand \@sanitize@url [0]{\catcode `\\12\catcode `\$12\catcode
  `\&12\catcode `\#12\catcode `\^12\catcode `\_12\catcode `\%12\relax}%
\providecommand \@@startlink[1]{}%
\providecommand \@@endlink[0]{}%
\providecommand \url  [0]{\begingroup\@sanitize@url \@url }%
\providecommand \@url [1]{\endgroup\@href {#1}{\urlprefix }}%
\providecommand \urlprefix  [0]{URL }%
\providecommand \Eprint [0]{\href }%
\providecommand \doibase [0]{http://dx.doi.org/}%
\providecommand \selectlanguage [0]{\@gobble}%
\providecommand \bibinfo  [0]{\@secondoftwo}%
\providecommand \bibfield  [0]{\@secondoftwo}%
\providecommand \translation [1]{[#1]}%
\providecommand \BibitemOpen [0]{}%
\providecommand \bibitemStop [0]{}%
\providecommand \bibitemNoStop [0]{.\EOS\space}%
\providecommand \EOS [0]{\spacefactor3000\relax}%
\providecommand \BibitemShut  [1]{\csname bibitem#1\endcsname}%
\let\auto@bib@innerbib\@empty
\bibitem [{\citenamefont {Sekatskii}\ and\ \citenamefont
  {Letokhov}(1996)}]{Sekatskii1996}%
  \BibitemOpen
  \bibfield  {author} {\bibinfo {author} {\bibfnamefont {S.}~\bibnamefont
  {Sekatskii}}\ and\ \bibinfo {author} {\bibfnamefont {V.}~\bibnamefont
  {Letokhov}},\ }\href {http://dx.doi.org/10.1134/1.567024} {\bibfield
  {journal} {\bibinfo  {journal} {JETP Letters}\ }\textbf {\bibinfo {volume}
  {63}},\ \bibinfo {pages} {319} (\bibinfo {year} {1996})}\BibitemShut
  {NoStop}%
\bibitem [{\citenamefont {Michaelis}\ \emph {et~al.}(2000)\citenamefont
  {Michaelis}, \citenamefont {Hettich}, \citenamefont {Mlynek},\ and\
  \citenamefont {Sandoghdar}}]{Michaelis2000}%
  \BibitemOpen
  \bibfield  {author} {\bibinfo {author} {\bibfnamefont {J.}~\bibnamefont
  {Michaelis}}, \bibinfo {author} {\bibfnamefont {C.}~\bibnamefont {Hettich}},
  \bibinfo {author} {\bibfnamefont {J.}~\bibnamefont {Mlynek}}, \ and\ \bibinfo
  {author} {\bibfnamefont {V.}~\bibnamefont {Sandoghdar}},\ }\href
  {http://dx.doi.org/10.1038/35012545} {\bibfield  {journal} {\bibinfo
  {journal} {Nature}\ }\textbf {\bibinfo {volume} {405}},\ \bibinfo {pages}
  {325} (\bibinfo {year} {2000})}\BibitemShut {NoStop}%
\bibitem [{\citenamefont {Chernobrod}\ and\ \citenamefont
  {Berman}(2005)}]{Chernobrod2005}%
  \BibitemOpen
  \bibfield  {author} {\bibinfo {author} {\bibfnamefont {B.~M.}\ \bibnamefont
  {Chernobrod}}\ and\ \bibinfo {author} {\bibfnamefont {G.~P.}\ \bibnamefont
  {Berman}},\ }\href {\doibase 10.1063/1.1829373} {\bibfield  {journal}
  {\bibinfo  {journal} {J. Appl. Phys.}\ }\textbf {\bibinfo {volume} {97}},\
  \bibinfo {pages} {014903} (\bibinfo {year} {2005})}\BibitemShut {NoStop}%
\bibitem [{\citenamefont {Balasubramanian}\ \emph {et~al.}(2008)\citenamefont
  {Balasubramanian}, \citenamefont {Chan}, \citenamefont {Kolesov},
  \citenamefont {Al-Hmoud}, \citenamefont {Tisler}, \citenamefont {Shin},
  \citenamefont {Kim}, \citenamefont {Wojcik}, \citenamefont {Hemmer},
  \citenamefont {Hanke}, \citenamefont {Leitenstorfer}, \citenamefont
  {Bratschitsch}, \citenamefont {Jelezko},\ and\ \citenamefont
  {Wrachtrup}}]{Balasubramanian2008}%
  \BibitemOpen
  \bibfield  {author} {\bibinfo {author} {\bibfnamefont {G.}~\bibnamefont
  {Balasubramanian}}, \bibinfo {author} {\bibfnamefont {I.}~\bibnamefont
  {Chan}}, \bibinfo {author} {\bibfnamefont {R.}~\bibnamefont {Kolesov}},
  \bibinfo {author} {\bibfnamefont {M.}~\bibnamefont {Al-Hmoud}}, \bibinfo
  {author} {\bibfnamefont {J.}~\bibnamefont {Tisler}}, \bibinfo {author}
  {\bibfnamefont {C.}~\bibnamefont {Shin}}, \bibinfo {author} {\bibfnamefont
  {C.}~\bibnamefont {Kim}}, \bibinfo {author} {\bibfnamefont {A.}~\bibnamefont
  {Wojcik}}, \bibinfo {author} {\bibfnamefont {A.}~\bibnamefont {Hemmer},
  \bibfnamefont {P.R. amd~Krueger}}, \bibinfo {author} {\bibfnamefont
  {T.}~\bibnamefont {Hanke}}, \bibinfo {author} {\bibfnamefont
  {A.}~\bibnamefont {Leitenstorfer}}, \bibinfo {author} {\bibfnamefont
  {R.}~\bibnamefont {Bratschitsch}}, \bibinfo {author} {\bibfnamefont
  {F.}~\bibnamefont {Jelezko}}, \ and\ \bibinfo {author} {\bibfnamefont
  {J.}~\bibnamefont {Wrachtrup}},\ }\href
  {http://dx.doi.org/10.1038/nature07278} {\bibfield  {journal} {\bibinfo
  {journal} {Nature}\ }\textbf {\bibinfo {volume} {455}},\ \bibinfo {pages}
  {606} (\bibinfo {year} {2008})}\BibitemShut {NoStop}%
\bibitem [{\citenamefont {Maze}\ \emph {et~al.}(2008)\citenamefont {Maze},
  \citenamefont {Stanwix}, \citenamefont {Hodges}, \citenamefont {Hong},
  \citenamefont {Taylor}, \citenamefont {Cappellaro}, \citenamefont {Jiang},
  \citenamefont {Gurudev~Dutt}, \citenamefont {Togan}, \citenamefont {Zibrov},
  \citenamefont {Yacoby}, \citenamefont {Walsworth},\ and\ \citenamefont
  {Lukin}}]{Maze2008a}%
  \BibitemOpen
  \bibfield  {author} {\bibinfo {author} {\bibfnamefont {J.~R.}\ \bibnamefont
  {Maze}}, \bibinfo {author} {\bibfnamefont {P.~L.}\ \bibnamefont {Stanwix}},
  \bibinfo {author} {\bibfnamefont {J.~S.}\ \bibnamefont {Hodges}}, \bibinfo
  {author} {\bibfnamefont {S.}~\bibnamefont {Hong}}, \bibinfo {author}
  {\bibfnamefont {J.~M.}\ \bibnamefont {Taylor}}, \bibinfo {author}
  {\bibfnamefont {P.}~\bibnamefont {Cappellaro}}, \bibinfo {author}
  {\bibfnamefont {L.}~\bibnamefont {Jiang}}, \bibinfo {author} {\bibfnamefont
  {M.~V.}\ \bibnamefont {Gurudev~Dutt}}, \bibinfo {author} {\bibfnamefont
  {E.}~\bibnamefont {Togan}}, \bibinfo {author} {\bibfnamefont {A.~S.}\
  \bibnamefont {Zibrov}}, \bibinfo {author} {\bibfnamefont {A.}~\bibnamefont
  {Yacoby}}, \bibinfo {author} {\bibfnamefont {R.~L.}\ \bibnamefont
  {Walsworth}}, \ and\ \bibinfo {author} {\bibfnamefont {M.~D.}\ \bibnamefont
  {Lukin}},\ }\href {http://dx.doi.org/10.1038/nature07279} {\bibfield
  {journal} {\bibinfo  {journal} {Nature}\ }\textbf {\bibinfo {volume} {455}},\
  \bibinfo {pages} {644} (\bibinfo {year} {2008})}\BibitemShut {NoStop}%
\bibitem [{\citenamefont {Taylor}\ \emph {et~al.}(2008)\citenamefont {Taylor},
  \citenamefont {Cappellaro}, \citenamefont {Childress}, \citenamefont {Jiang},
  \citenamefont {Budker}, \citenamefont {Hemmer}, \citenamefont {Yacoby},
  \citenamefont {Walsworth},\ and\ \citenamefont {Lukin}}]{Taylor2008}%
  \BibitemOpen
  \bibfield  {author} {\bibinfo {author} {\bibfnamefont {J.}~\bibnamefont
  {Taylor}}, \bibinfo {author} {\bibfnamefont {P.}~\bibnamefont {Cappellaro}},
  \bibinfo {author} {\bibfnamefont {L.}~\bibnamefont {Childress}}, \bibinfo
  {author} {\bibfnamefont {L.}~\bibnamefont {Jiang}}, \bibinfo {author}
  {\bibfnamefont {D.}~\bibnamefont {Budker}}, \bibinfo {author} {\bibfnamefont
  {P.}~\bibnamefont {Hemmer}}, \bibinfo {author} {\bibfnamefont
  {A.}~\bibnamefont {Yacoby}}, \bibinfo {author} {\bibfnamefont
  {R.}~\bibnamefont {Walsworth}}, \ and\ \bibinfo {author} {\bibfnamefont
  {M.}~\bibnamefont {Lukin}},\ }\href {http://dx.doi.org/10.1038/nphys1075}
  {\bibfield  {journal} {\bibinfo  {journal} {Nature Physics}\ }\textbf
  {\bibinfo {volume} {4}},\ \bibinfo {pages} {810} (\bibinfo {year}
  {2008})}\BibitemShut {NoStop}%
\bibitem [{\citenamefont {Dolde}\ \emph {et~al.}(2011)\citenamefont {Dolde},
  \citenamefont {Fedder}, \citenamefont {Doherty}, \citenamefont {N\"obauer},
  \citenamefont {Rempp}, \citenamefont {Balasubramanian}, \citenamefont {Wolf},
  \citenamefont {Reinhard}, \citenamefont {Hollenberg}, \citenamefont
  {Jelezko},\ and\ \citenamefont {Wrachtrup}}]{Dolde2011}%
  \BibitemOpen
  \bibfield  {author} {\bibinfo {author} {\bibfnamefont {F.}~\bibnamefont
  {Dolde}}, \bibinfo {author} {\bibfnamefont {H.}~\bibnamefont {Fedder}},
  \bibinfo {author} {\bibfnamefont {M.~W.}\ \bibnamefont {Doherty}}, \bibinfo
  {author} {\bibfnamefont {T.}~\bibnamefont {N\"obauer}}, \bibinfo {author}
  {\bibfnamefont {F.}~\bibnamefont {Rempp}}, \bibinfo {author} {\bibfnamefont
  {G.}~\bibnamefont {Balasubramanian}}, \bibinfo {author} {\bibfnamefont
  {T.}~\bibnamefont {Wolf}}, \bibinfo {author} {\bibfnamefont {F.}~\bibnamefont
  {Reinhard}}, \bibinfo {author} {\bibfnamefont {L.~C.~L.}\ \bibnamefont
  {Hollenberg}}, \bibinfo {author} {\bibfnamefont {F.}~\bibnamefont {Jelezko}},
  \ and\ \bibinfo {author} {\bibfnamefont {J.}~\bibnamefont {Wrachtrup}},\
  }\href {http://dx.doi.org/10.1038/nphys1969} {\bibfield  {journal} {\bibinfo
  {journal} {Nature Physics}\ }\textbf {\bibinfo {volume} {7}},\ \bibinfo
  {pages} {459} (\bibinfo {year} {2011})}\BibitemShut {NoStop}%
\bibitem [{\citenamefont {Degen}(2008)}]{Degen2008}%
  \BibitemOpen
  \bibfield  {author} {\bibinfo {author} {\bibfnamefont {C.~L.}\ \bibnamefont
  {Degen}},\ }\href {http://dx.doi.org/10.1063/1.2943282} {\bibfield  {journal}
  {\bibinfo  {journal} {Appl. Phys. Lett.}\ }\textbf {\bibinfo {volume} {92}},\
  \bibinfo {eid} {243111} (\bibinfo {year} {2008})}\BibitemShut {NoStop}%
\bibitem [{\citenamefont {Neumann}\ \emph {et~al.}(2010)\citenamefont
  {Neumann}, \citenamefont {Kolesov}, \citenamefont {Naydenov}, \citenamefont
  {Beck}, \citenamefont {Rempp}, \citenamefont {Steiner}, \citenamefont
  {Jacques}, \citenamefont {Balasubramanian}, \citenamefont {Markham},
  \citenamefont {Twitchen}, \citenamefont {Pezzagna}, \citenamefont {Meijer},
  \citenamefont {Twamley}, \citenamefont {Jelezko},\ and\ \citenamefont
  {Wrachtrup}}]{Neumann2010}%
  \BibitemOpen
  \bibfield  {author} {\bibinfo {author} {\bibfnamefont {P.}~\bibnamefont
  {Neumann}}, \bibinfo {author} {\bibfnamefont {R.}~\bibnamefont {Kolesov}},
  \bibinfo {author} {\bibfnamefont {B.}~\bibnamefont {Naydenov}}, \bibinfo
  {author} {\bibfnamefont {J.}~\bibnamefont {Beck}}, \bibinfo {author}
  {\bibfnamefont {F.}~\bibnamefont {Rempp}}, \bibinfo {author} {\bibfnamefont
  {M.}~\bibnamefont {Steiner}}, \bibinfo {author} {\bibfnamefont
  {V.}~\bibnamefont {Jacques}}, \bibinfo {author} {\bibfnamefont
  {G.}~\bibnamefont {Balasubramanian}}, \bibinfo {author} {\bibfnamefont
  {M.~L.}\ \bibnamefont {Markham}}, \bibinfo {author} {\bibfnamefont {D.~J.}\
  \bibnamefont {Twitchen}}, \bibinfo {author} {\bibfnamefont {S.}~\bibnamefont
  {Pezzagna}}, \bibinfo {author} {\bibfnamefont {J.}~\bibnamefont {Meijer}},
  \bibinfo {author} {\bibfnamefont {J.}~\bibnamefont {Twamley}}, \bibinfo
  {author} {\bibfnamefont {F.}~\bibnamefont {Jelezko}}, \ and\ \bibinfo
  {author} {\bibfnamefont {J.}~\bibnamefont {Wrachtrup}},\ }\href
  {http://dx.doi.org/10.1038/nphys1536} {\bibfield  {journal} {\bibinfo
  {journal} {Nat Phys}\ }\textbf {\bibinfo {volume} {6}},\ \bibinfo {pages}
  {249} (\bibinfo {year} {2010})}\BibitemShut {NoStop}%
\bibitem [{\citenamefont {McGuinness}\ \emph {et~al.}(2011)\citenamefont
  {McGuinness}, \citenamefont {Yan}, \citenamefont {Stacey}, \citenamefont
  {Simpson}, \citenamefont {Hall}, \citenamefont {Maclaurin}, \citenamefont
  {Prawer}, \citenamefont {Mulvaney}, \citenamefont {Wrachtrup}, \citenamefont
  {Caruso}, \citenamefont {Scholten},\ and\ \citenamefont
  {Hollenberg}}]{McGuinness2011}%
  \BibitemOpen
  \bibfield  {author} {\bibinfo {author} {\bibfnamefont {L.~P.}\ \bibnamefont
  {McGuinness}}, \bibinfo {author} {\bibfnamefont {Y.}~\bibnamefont {Yan}},
  \bibinfo {author} {\bibfnamefont {A.}~\bibnamefont {Stacey}}, \bibinfo
  {author} {\bibfnamefont {D.~A.}\ \bibnamefont {Simpson}}, \bibinfo {author}
  {\bibfnamefont {L.~T.}\ \bibnamefont {Hall}}, \bibinfo {author}
  {\bibfnamefont {D.}~\bibnamefont {Maclaurin}}, \bibinfo {author}
  {\bibfnamefont {S.}~\bibnamefont {Prawer}}, \bibinfo {author} {\bibfnamefont
  {P.}~\bibnamefont {Mulvaney}}, \bibinfo {author} {\bibfnamefont
  {J.}~\bibnamefont {Wrachtrup}}, \bibinfo {author} {\bibfnamefont
  {F.}~\bibnamefont {Caruso}}, \bibinfo {author} {\bibfnamefont {R.~E.}\
  \bibnamefont {Scholten}}, \ and\ \bibinfo {author} {\bibfnamefont {L.~C.~L.}\
  \bibnamefont {Hollenberg}},\ }\href {http://dx.doi.org/10.1038/nnano.2011.64}
  {\bibfield  {journal} {\bibinfo  {journal} {Nat Nano}\ }\textbf {\bibinfo
  {volume} {6}},\ \bibinfo {pages} {358} (\bibinfo {year} {2011})}\BibitemShut
  {NoStop}%
\bibitem [{\citenamefont {Kuhn}\ \emph {et~al.}(2001)\citenamefont {Kuhn},
  \citenamefont {Hettich}, \citenamefont {Schmitt}, \citenamefont {Poizat},\
  and\ \citenamefont {Sandoghdar}}]{Kuhn2001}%
  \BibitemOpen
  \bibfield  {author} {\bibinfo {author} {\bibfnamefont {S.}~\bibnamefont
  {Kuhn}}, \bibinfo {author} {\bibfnamefont {C.}~\bibnamefont {Hettich}},
  \bibinfo {author} {\bibfnamefont {C.}~\bibnamefont {Schmitt}}, \bibinfo
  {author} {\bibfnamefont {J.}~\bibnamefont {Poizat}}, \ and\ \bibinfo {author}
  {\bibfnamefont {V.}~\bibnamefont {Sandoghdar}},\ }\href
  {http://dx.doi.org/10.1046/j.1365-2818.2001.00829.x} {\bibfield  {journal}
  {\bibinfo  {journal} {Journal of Microscopy}\ }\textbf {\bibinfo {volume}
  {202}},\ \bibinfo {pages} {2} (\bibinfo {year} {2001})}\BibitemShut {NoStop}%
\bibitem [{\citenamefont {Cuche}\ \emph {et~al.}(2010)\citenamefont {Cuche},
  \citenamefont {Drezet}, \citenamefont {Roch}, \citenamefont {Treussart},\
  and\ \citenamefont {Huant}}]{Cuche2010}%
  \BibitemOpen
  \bibfield  {author} {\bibinfo {author} {\bibfnamefont {A.}~\bibnamefont
  {Cuche}}, \bibinfo {author} {\bibfnamefont {A.}~\bibnamefont {Drezet}},
  \bibinfo {author} {\bibfnamefont {J.-F.}\ \bibnamefont {Roch}}, \bibinfo
  {author} {\bibfnamefont {F.}~\bibnamefont {Treussart}}, \ and\ \bibinfo
  {author} {\bibfnamefont {S.}~\bibnamefont {Huant}},\ }\href
  {http://dx.doi.org/10.1117/1.3374237} {\bibfield  {journal} {\bibinfo
  {journal} {Journal of Nanophotonics}\ }\textbf {\bibinfo {volume} {4}},\
  \bibinfo {eid} {043506} (\bibinfo {year} {2010})}\BibitemShut {NoStop}%
\bibitem [{\citenamefont {Kurtsiefer}\ \emph {et~al.}(2000)\citenamefont
  {Kurtsiefer}, \citenamefont {Mayer}, \citenamefont {Zarda},\ and\
  \citenamefont {Weinfurter}}]{Kurtsiefer2000}%
  \BibitemOpen
  \bibfield  {author} {\bibinfo {author} {\bibfnamefont {C.}~\bibnamefont
  {Kurtsiefer}}, \bibinfo {author} {\bibfnamefont {S.}~\bibnamefont {Mayer}},
  \bibinfo {author} {\bibfnamefont {P.}~\bibnamefont {Zarda}}, \ and\ \bibinfo
  {author} {\bibfnamefont {H.}~\bibnamefont {Weinfurter}},\ }\href
  {http://dx.doi.org/10.1103/PhysRevLett.85.290} {\bibfield  {journal}
  {\bibinfo  {journal} {Phys. Rev. Lett.}\ }\textbf {\bibinfo {volume} {85}},\
  \bibinfo {pages} {290} (\bibinfo {year} {2000})}\BibitemShut {NoStop}%
\bibitem [{\citenamefont {Balasubramanian}\ \emph {et~al.}(2009)\citenamefont
  {Balasubramanian}, \citenamefont {Neumann}, \citenamefont {Twitchen},
  \citenamefont {Markham}, \citenamefont {Kolesov}, \citenamefont {Mizuochi},
  \citenamefont {Isoya}, \citenamefont {Achard}, \citenamefont {Beck},
  \citenamefont {Tissler}, \citenamefont {Jacques}, \citenamefont {Hemmer},
  \citenamefont {Jelezko},\ and\ \citenamefont
  {Wrachtrup}}]{Balasubramanian2009}%
  \BibitemOpen
  \bibfield  {author} {\bibinfo {author} {\bibfnamefont {G.}~\bibnamefont
  {Balasubramanian}}, \bibinfo {author} {\bibfnamefont {P.}~\bibnamefont
  {Neumann}}, \bibinfo {author} {\bibfnamefont {D.}~\bibnamefont {Twitchen}},
  \bibinfo {author} {\bibfnamefont {M.}~\bibnamefont {Markham}}, \bibinfo
  {author} {\bibfnamefont {R.}~\bibnamefont {Kolesov}}, \bibinfo {author}
  {\bibfnamefont {N.}~\bibnamefont {Mizuochi}}, \bibinfo {author}
  {\bibfnamefont {J.}~\bibnamefont {Isoya}}, \bibinfo {author} {\bibfnamefont
  {J.}~\bibnamefont {Achard}}, \bibinfo {author} {\bibfnamefont
  {J.}~\bibnamefont {Beck}}, \bibinfo {author} {\bibfnamefont {J.}~\bibnamefont
  {Tissler}}, \bibinfo {author} {\bibfnamefont {V.}~\bibnamefont {Jacques}},
  \bibinfo {author} {\bibfnamefont {P.}~\bibnamefont {Hemmer}}, \bibinfo
  {author} {\bibfnamefont {F.}~\bibnamefont {Jelezko}}, \ and\ \bibinfo
  {author} {\bibfnamefont {J.}~\bibnamefont {Wrachtrup}},\ }\href
  {http://dx.doi.org/10.1038/nmat2420} {\bibfield  {journal} {\bibinfo
  {journal} {Nature Materials}\ }\textbf {\bibinfo {volume} {8}},\ \bibinfo
  {pages} {383} (\bibinfo {year} {2009})}\BibitemShut {NoStop}%
\bibitem [{\citenamefont {Gruber}\ \emph {et~al.}(1997)\citenamefont {Gruber},
  \citenamefont {Drabenstedt}, \citenamefont {Tietz}, \citenamefont {Fleury},
  \citenamefont {Wrachtrup},\ and\ \citenamefont {Borczyskowski}}]{Gruber1997}%
  \BibitemOpen
  \bibfield  {author} {\bibinfo {author} {\bibfnamefont {A.}~\bibnamefont
  {Gruber}}, \bibinfo {author} {\bibfnamefont {A.}~\bibnamefont {Drabenstedt}},
  \bibinfo {author} {\bibfnamefont {C.}~\bibnamefont {Tietz}}, \bibinfo
  {author} {\bibfnamefont {L.}~\bibnamefont {Fleury}}, \bibinfo {author}
  {\bibfnamefont {J.}~\bibnamefont {Wrachtrup}}, \ and\ \bibinfo {author}
  {\bibfnamefont {C.}~\bibnamefont {Borczyskowski}},\ }\href
  {http://dx.doi.org/10.1126/science.276.5321.2012} {\bibfield  {journal}
  {\bibinfo  {journal} {Science}\ }\textbf {\bibinfo {volume} {276}},\ \bibinfo
  {pages} {2012} (\bibinfo {year} {1997})}\BibitemShut {NoStop}%
\bibitem [{\citenamefont {Bradac}\ \emph {et~al.}(2010)\citenamefont {Bradac},
  \citenamefont {Gaebel}, \citenamefont {Naidoo}, \citenamefont {Sellars},
  \citenamefont {Twamley}, \citenamefont {Brown}, \citenamefont {Barnard},
  \citenamefont {Plakhotnik}, \citenamefont {Zvyagin},\ and\ \citenamefont
  {Rabeau}}]{Bradac2010}%
  \BibitemOpen
  \bibfield  {author} {\bibinfo {author} {\bibfnamefont {C.}~\bibnamefont
  {Bradac}}, \bibinfo {author} {\bibfnamefont {T.}~\bibnamefont {Gaebel}},
  \bibinfo {author} {\bibfnamefont {N.}~\bibnamefont {Naidoo}}, \bibinfo
  {author} {\bibfnamefont {M.~J.}\ \bibnamefont {Sellars}}, \bibinfo {author}
  {\bibfnamefont {J.}~\bibnamefont {Twamley}}, \bibinfo {author} {\bibfnamefont
  {L.~J.}\ \bibnamefont {Brown}}, \bibinfo {author} {\bibfnamefont {A.~S.}\
  \bibnamefont {Barnard}}, \bibinfo {author} {\bibfnamefont {T.}~\bibnamefont
  {Plakhotnik}}, \bibinfo {author} {\bibfnamefont {A.~V.}\ \bibnamefont
  {Zvyagin}}, \ and\ \bibinfo {author} {\bibfnamefont {J.~R.}\ \bibnamefont
  {Rabeau}},\ }\href {http://dx.doi.org/10.1038/nnano.2010.56} {\bibfield
  {journal} {\bibinfo  {journal} {Nature Nanotechnology}\ }\textbf {\bibinfo
  {volume} {5}},\ \bibinfo {pages} {345} (\bibinfo {year} {2010})}\BibitemShut
  {NoStop}%
\bibitem [{\citenamefont {Stanwix}\ \emph {et~al.}(2010)\citenamefont
  {Stanwix}, \citenamefont {Pham}, \citenamefont {Maze}, \citenamefont
  {Le~Sage}, \citenamefont {Yeung}, \citenamefont {Cappellaro}, \citenamefont
  {Hemmer}, \citenamefont {Yacoby}, \citenamefont {Lukin},\ and\ \citenamefont
  {Walsworth}}]{Stanwix2010}%
  \BibitemOpen
  \bibfield  {author} {\bibinfo {author} {\bibfnamefont {P.~L.}\ \bibnamefont
  {Stanwix}}, \bibinfo {author} {\bibfnamefont {L.~M.}\ \bibnamefont {Pham}},
  \bibinfo {author} {\bibfnamefont {J.~R.}\ \bibnamefont {Maze}}, \bibinfo
  {author} {\bibfnamefont {D.}~\bibnamefont {Le~Sage}}, \bibinfo {author}
  {\bibfnamefont {T.~K.}\ \bibnamefont {Yeung}}, \bibinfo {author}
  {\bibfnamefont {P.}~\bibnamefont {Cappellaro}}, \bibinfo {author}
  {\bibfnamefont {P.~R.}\ \bibnamefont {Hemmer}}, \bibinfo {author}
  {\bibfnamefont {A.}~\bibnamefont {Yacoby}}, \bibinfo {author} {\bibfnamefont
  {M.~D.}\ \bibnamefont {Lukin}}, \ and\ \bibinfo {author} {\bibfnamefont
  {R.~L.}\ \bibnamefont {Walsworth}},\ }\href
  {http://dx.doi.org/10.1103/PhysRevB.82.201201} {\bibfield  {journal}
  {\bibinfo  {journal} {Phys. Rev. B}\ }\textbf {\bibinfo {volume} {82}},\
  \bibinfo {pages} {201201} (\bibinfo {year} {2010})}\BibitemShut {NoStop}%
\bibitem [{\citenamefont {Babinec}\ \emph {et~al.}(2010)\citenamefont
  {Babinec}, \citenamefont {Hausmann}, \citenamefont {Khan}, \citenamefont
  {Zhang}, \citenamefont {Maze}, \citenamefont {Hemmer},\ and\ \citenamefont
  {Loncar}}]{Babinec2010}%
  \BibitemOpen
  \bibfield  {author} {\bibinfo {author} {\bibfnamefont {T.~M.}\ \bibnamefont
  {Babinec}}, \bibinfo {author} {\bibfnamefont {B.~J.~M.}\ \bibnamefont
  {Hausmann}}, \bibinfo {author} {\bibfnamefont {M.}~\bibnamefont {Khan}},
  \bibinfo {author} {\bibfnamefont {Y.}~\bibnamefont {Zhang}}, \bibinfo
  {author} {\bibfnamefont {J.~R.}\ \bibnamefont {Maze}}, \bibinfo {author}
  {\bibfnamefont {P.~R.}\ \bibnamefont {Hemmer}}, \ and\ \bibinfo {author}
  {\bibfnamefont {M.}~\bibnamefont {Loncar}},\ }\href
  {http://dx.doi.org/10.1038/nnano.2010.6} {\bibfield  {journal} {\bibinfo
  {journal} {Nature Nanotechnology}\ }\textbf {\bibinfo {volume} {5}},\
  \bibinfo {pages} {195} (\bibinfo {year} {2010})}\BibitemShut {NoStop}%
\bibitem [{\citenamefont {Hausmann}\ \emph {et~al.}(2010)\citenamefont
  {Hausmann}, \citenamefont {Khan}, \citenamefont {Zhang}, \citenamefont
  {Babinec}, \citenamefont {Martinick}, \citenamefont {McCutcheon},
  \citenamefont {Hemmer},\ and\ \citenamefont {Loncar}}]{Hausmann2010}%
  \BibitemOpen
  \bibfield  {author} {\bibinfo {author} {\bibfnamefont {B.~J.}\ \bibnamefont
  {Hausmann}}, \bibinfo {author} {\bibfnamefont {M.}~\bibnamefont {Khan}},
  \bibinfo {author} {\bibfnamefont {Y.}~\bibnamefont {Zhang}}, \bibinfo
  {author} {\bibfnamefont {T.~M.}\ \bibnamefont {Babinec}}, \bibinfo {author}
  {\bibfnamefont {K.}~\bibnamefont {Martinick}}, \bibinfo {author}
  {\bibfnamefont {M.}~\bibnamefont {McCutcheon}}, \bibinfo {author}
  {\bibfnamefont {P.~R.}\ \bibnamefont {Hemmer}}, \ and\ \bibinfo {author}
  {\bibfnamefont {M.}~\bibnamefont {Loncar}},\ }\href {\doibase DOI:
  10.1016/j.diamond.2010.01.011} {\bibfield  {journal} {\bibinfo  {journal}
  {Diamond and Related Materials}\ }\textbf {\bibinfo {volume} {19}},\ \bibinfo
  {pages} {621 } (\bibinfo {year} {2010})},\ \bibinfo {note} {proceedings of
  Diamond 2009, The 20th European Conference on Diamond, Diamond-Like
  Materials, Carbon Nanotubes and Nitrides, Part 1}\BibitemShut {NoStop}%
\bibitem [{\citenamefont {Kalish}\ \emph {et~al.}(1997)\citenamefont {Kalish},
  \citenamefont {Uzan-Saguy}, \citenamefont {Philosoph}, \citenamefont
  {Richter}, \citenamefont {Lagrange}, \citenamefont {Gheeraert}, \citenamefont
  {Deneuville},\ and\ \citenamefont {Collins}}]{Kalish1997}%
  \BibitemOpen
  \bibfield  {author} {\bibinfo {author} {\bibfnamefont {R.}~\bibnamefont
  {Kalish}}, \bibinfo {author} {\bibfnamefont {C.}~\bibnamefont {Uzan-Saguy}},
  \bibinfo {author} {\bibfnamefont {B.}~\bibnamefont {Philosoph}}, \bibinfo
  {author} {\bibfnamefont {V.}~\bibnamefont {Richter}}, \bibinfo {author}
  {\bibfnamefont {J.~P.}\ \bibnamefont {Lagrange}}, \bibinfo {author}
  {\bibfnamefont {E.}~\bibnamefont {Gheeraert}}, \bibinfo {author}
  {\bibfnamefont {A.}~\bibnamefont {Deneuville}}, \ and\ \bibinfo {author}
  {\bibfnamefont {A.~T.}\ \bibnamefont {Collins}},\ }\href {\doibase DOI:
  10.1016/S0925-9635(96)00657-7} {\bibfield  {journal} {\bibinfo  {journal}
  {Diamond and Related Materials}\ }\textbf {\bibinfo {volume} {6}},\ \bibinfo
  {pages} {516 } (\bibinfo {year} {1997})}\BibitemShut {NoStop}%
\bibitem [{\citenamefont {Jelezko}\ \emph {et~al.}(2004)\citenamefont
  {Jelezko}, \citenamefont {Gaebel}, \citenamefont {Popa}, \citenamefont
  {Gruber},\ and\ \citenamefont {Wrachtrup}}]{Jelezko2004}%
  \BibitemOpen
  \bibfield  {author} {\bibinfo {author} {\bibfnamefont {F.}~\bibnamefont
  {Jelezko}}, \bibinfo {author} {\bibfnamefont {T.}~\bibnamefont {Gaebel}},
  \bibinfo {author} {\bibfnamefont {I.}~\bibnamefont {Popa}}, \bibinfo {author}
  {\bibfnamefont {A.}~\bibnamefont {Gruber}}, \ and\ \bibinfo {author}
  {\bibfnamefont {J.}~\bibnamefont {Wrachtrup}},\ }\href
  {http://dx.doi.org/10.1103/PhysRevLett.92.076401} {\bibfield  {journal}
  {\bibinfo  {journal} {Phys. Rev. Lett.}\ }\textbf {\bibinfo {volume} {92}},\
  \bibinfo {pages} {076401} (\bibinfo {year} {2004})}\BibitemShut {NoStop}%
\bibitem [{\citenamefont {Oort}\ and\ \citenamefont
  {Glasbeek}(1990)}]{VanOort1990}%
  \BibitemOpen
  \bibfield  {author} {\bibinfo {author} {\bibfnamefont {E.~V.}\ \bibnamefont
  {Oort}}\ and\ \bibinfo {author} {\bibfnamefont {M.}~\bibnamefont
  {Glasbeek}},\ }\href {http://dx.doi.org/10.1016/0301-0104(90)85013-M}
  {\bibfield  {journal} {\bibinfo  {journal} {Chemical Physics}\ }\textbf
  {\bibinfo {volume} {143}},\ \bibinfo {pages} {131} (\bibinfo {year}
  {1990})}\BibitemShut {NoStop}%
\bibitem [{\citenamefont {Grinolds}\ \emph {et~al.}(2011)\citenamefont
  {Grinolds}, \citenamefont {Maletinsky}, \citenamefont {Hong}, \citenamefont
  {Lukin}, \citenamefont {Walsworth},\ and\ \citenamefont
  {Yacoby}}]{Grinolds2011}%
  \BibitemOpen
  \bibfield  {author} {\bibinfo {author} {\bibfnamefont {M.~S.}\ \bibnamefont
  {Grinolds}}, \bibinfo {author} {\bibfnamefont {P.}~\bibnamefont
  {Maletinsky}}, \bibinfo {author} {\bibfnamefont {S.}~\bibnamefont {Hong}},
  \bibinfo {author} {\bibfnamefont {M.~D.}\ \bibnamefont {Lukin}}, \bibinfo
  {author} {\bibfnamefont {R.~L.}\ \bibnamefont {Walsworth}}, \ and\ \bibinfo
  {author} {\bibfnamefont {A.}~\bibnamefont {Yacoby}},\ }\href
  {http://dx.doi.org/10.1038/nphys1999} {\bibfield  {journal} {\bibinfo
  {journal} {Nature Physics}\ }\textbf {\bibinfo {volume} {0}},\ \bibinfo
  {pages} {0} (\bibinfo {year} {2011})}\BibitemShut {NoStop}%
\bibitem [{\citenamefont {Buchler}\ \emph {et~al.}(2005)\citenamefont
  {Buchler}, \citenamefont {Kalkbrenner}, \citenamefont {Hettich},\ and\
  \citenamefont {Sandoghdar}}]{Buchler2005}%
  \BibitemOpen
  \bibfield  {author} {\bibinfo {author} {\bibfnamefont {B.~C.}\ \bibnamefont
  {Buchler}}, \bibinfo {author} {\bibfnamefont {T.}~\bibnamefont
  {Kalkbrenner}}, \bibinfo {author} {\bibfnamefont {C.}~\bibnamefont
  {Hettich}}, \ and\ \bibinfo {author} {\bibfnamefont {V.}~\bibnamefont
  {Sandoghdar}},\ }\href {http://dx.doi.org/10.1103/PhysRevLett.95.063003}
  {\bibfield  {journal} {\bibinfo  {journal} {Phys. Rev. Lett.}\ }\textbf
  {\bibinfo {volume} {95}},\ \bibinfo {pages} {063003} (\bibinfo {year}
  {2005})}\BibitemShut {NoStop}%
\bibitem [{\citenamefont {Wolny}\ \emph {et~al.}(2010)\citenamefont {Wolny},
  \citenamefont {M\"uhl}, \citenamefont {Weissker}, \citenamefont {Lipert},
  \citenamefont {Schumann}, \citenamefont {Leonhardt},\ and\ \citenamefont
  {Büchner}}]{Wolny2010}%
  \BibitemOpen
  \bibfield  {author} {\bibinfo {author} {\bibfnamefont {F.}~\bibnamefont
  {Wolny}}, \bibinfo {author} {\bibfnamefont {T.}~\bibnamefont {M\"uhl}},
  \bibinfo {author} {\bibfnamefont {U.}~\bibnamefont {Weissker}}, \bibinfo
  {author} {\bibfnamefont {K.}~\bibnamefont {Lipert}}, \bibinfo {author}
  {\bibfnamefont {J.}~\bibnamefont {Schumann}}, \bibinfo {author}
  {\bibfnamefont {A.}~\bibnamefont {Leonhardt}}, \ and\ \bibinfo {author}
  {\bibfnamefont {B.}~\bibnamefont {Büchner}},\ }\href
  {http://stacks.iop.org/0957-4484/21/i=43/a=435501} {\bibfield  {journal}
  {\bibinfo  {journal} {Nanotechnology}\ }\textbf {\bibinfo {volume} {21}},\
  \bibinfo {pages} {435501} (\bibinfo {year} {2010})}\BibitemShut {NoStop}%
\bibitem [{\citenamefont {Kohashi}\ \emph {et~al.}(2009)\citenamefont
  {Kohashi}, \citenamefont {Konoto},\ and\ \citenamefont
  {Koike}}]{Kohashi2009}%
  \BibitemOpen
  \bibfield  {author} {\bibinfo {author} {\bibfnamefont {T.}~\bibnamefont
  {Kohashi}}, \bibinfo {author} {\bibfnamefont {M.}~\bibnamefont {Konoto}}, \
  and\ \bibinfo {author} {\bibfnamefont {K.}~\bibnamefont {Koike}},\ }\href
  {\doibase 10.1093/jmicro/dfp047} {\bibfield  {journal} {\bibinfo  {journal}
  {Journal of Electron Microscopy}\ }\textbf {\bibinfo {volume} {19}},\
  \bibinfo {pages} {1} (\bibinfo {year} {2009})}\BibitemShut {NoStop}%
\bibitem [{\citenamefont {Mollet}\ \emph {et~al.}(2011)\citenamefont {Mollet},
  \citenamefont {Cuche}, \citenamefont {Drezet},\ and\ \citenamefont
  {Huant}}]{Mollet2011}%
  \BibitemOpen
  \bibfield  {author} {\bibinfo {author} {\bibfnamefont {O.}~\bibnamefont
  {Mollet}}, \bibinfo {author} {\bibfnamefont {A.}~\bibnamefont {Cuche}},
  \bibinfo {author} {\bibfnamefont {A.}~\bibnamefont {Drezet}}, \ and\ \bibinfo
  {author} {\bibfnamefont {S.}~\bibnamefont {Huant}},\ }\href {\doibase DOI:
  10.1016/j.diamond.2011.05.012} {\bibfield  {journal} {\bibinfo  {journal}
  {Diamond and Related Materials}\ }\textbf {\bibinfo {volume} {20}},\ \bibinfo
  {pages} {995 } (\bibinfo {year} {2011})}\BibitemShut {NoStop}%
\bibitem [{\citenamefont {Kane}(1998)}]{Kane1998}%
  \BibitemOpen
  \bibfield  {author} {\bibinfo {author} {\bibfnamefont {B.~E.}\ \bibnamefont
  {Kane}},\ }\href {http://dx.doi.org/10.1038/30156} {\bibfield  {journal}
  {\bibinfo  {journal} {Nature}\ }\textbf {\bibinfo {volume} {393}},\ \bibinfo
  {pages} {133} (\bibinfo {year} {1998})}\BibitemShut {NoStop}%
\bibitem [{\citenamefont {Recher}\ and\ \citenamefont
  {Trauzettel}(2010)}]{Recher2010}%
  \BibitemOpen
  \bibfield  {author} {\bibinfo {author} {\bibfnamefont {P.}~\bibnamefont
  {Recher}}\ and\ \bibinfo {author} {\bibfnamefont {B.}~\bibnamefont
  {Trauzettel}},\ }\href {http://stacks.iop.org/0957-4484/21/i=30/a=302001}
  {\bibfield  {journal} {\bibinfo  {journal} {Nanotechnology}\ }\textbf
  {\bibinfo {volume} {21}},\ \bibinfo {pages} {302001} (\bibinfo {year}
  {2010})}\BibitemShut {NoStop}%
\bibitem [{\citenamefont {Togan}\ \emph {et~al.}(2010)\citenamefont {Togan},
  \citenamefont {Chu}, \citenamefont {Trifonov}, \citenamefont {Jiang},
  \citenamefont {Maze}, \citenamefont {Childress}, \citenamefont {Dutt},
  \citenamefont {Hemmer}, \citenamefont {Zibrov},\ and\ \citenamefont
  {Lukin}}]{Togan2010}%
  \BibitemOpen
  \bibfield  {author} {\bibinfo {author} {\bibfnamefont {E.}~\bibnamefont
  {Togan}}, \bibinfo {author} {\bibfnamefont {Y.}~\bibnamefont {Chu}}, \bibinfo
  {author} {\bibfnamefont {A.~S.}\ \bibnamefont {Trifonov}}, \bibinfo {author}
  {\bibfnamefont {L.}~\bibnamefont {Jiang}}, \bibinfo {author} {\bibfnamefont
  {J.}~\bibnamefont {Maze}}, \bibinfo {author} {\bibfnamefont {L.}~\bibnamefont
  {Childress}}, \bibinfo {author} {\bibfnamefont {A.~S.}\ \bibnamefont {Dutt},
  \bibfnamefont {M.~V. G. AU~Sorensen}}, \bibinfo {author} {\bibfnamefont
  {P.~R.}\ \bibnamefont {Hemmer}}, \bibinfo {author} {\bibfnamefont {A.~S.}\
  \bibnamefont {Zibrov}}, \ and\ \bibinfo {author} {\bibfnamefont {M.~D.}\
  \bibnamefont {Lukin}},\ }\href {http://dx.doi.org/10.1038/nature09256}
  {\bibfield  {journal} {\bibinfo  {journal} {Nature}\ }\textbf {\bibinfo
  {volume} {466}},\ \bibinfo {pages} {730} (\bibinfo {year}
  {2010})}\BibitemShut {NoStop}%
\bibitem [{\citenamefont {Lee}\ \emph {et~al.}(2008)\citenamefont {Lee},
  \citenamefont {Gu}, \citenamefont {Dawson}, \citenamefont {Friel},\ and\
  \citenamefont {Scarsbrook}}]{Lee2008}%
  \BibitemOpen
  \bibfield  {author} {\bibinfo {author} {\bibfnamefont {C.}~\bibnamefont
  {Lee}}, \bibinfo {author} {\bibfnamefont {E.}~\bibnamefont {Gu}}, \bibinfo
  {author} {\bibfnamefont {M.}~\bibnamefont {Dawson}}, \bibinfo {author}
  {\bibfnamefont {I.}~\bibnamefont {Friel}}, \ and\ \bibinfo {author}
  {\bibfnamefont {G.}~\bibnamefont {Scarsbrook}},\ }\href
  {http://dx.doi.org/10.1016/j.diamond.2008.01.011} {\bibfield  {journal}
  {\bibinfo  {journal} {Diamond and Related Materials}\ }\textbf {\bibinfo
  {volume} {17}},\ \bibinfo {pages} {1292} (\bibinfo {year}
  {2008})}\BibitemShut {NoStop}%
\bibitem [{\citenamefont {Childress}\ \emph {et~al.}(2006)\citenamefont
  {Childress}, \citenamefont {Gurudev~Dutt}, \citenamefont {Taylor},
  \citenamefont {Zibrov}, \citenamefont {Jelezko}, \citenamefont {Wrachtrup},
  \citenamefont {Hemmer},\ and\ \citenamefont {Lukin}}]{Childress2006}%
  \BibitemOpen
  \bibfield  {author} {\bibinfo {author} {\bibfnamefont {L.}~\bibnamefont
  {Childress}}, \bibinfo {author} {\bibfnamefont {M.}~\bibnamefont
  {Gurudev~Dutt}}, \bibinfo {author} {\bibfnamefont {J.}~\bibnamefont
  {Taylor}}, \bibinfo {author} {\bibfnamefont {A.}~\bibnamefont {Zibrov}},
  \bibinfo {author} {\bibfnamefont {F.}~\bibnamefont {Jelezko}}, \bibinfo
  {author} {\bibfnamefont {J.}~\bibnamefont {Wrachtrup}}, \bibinfo {author}
  {\bibfnamefont {P.}~\bibnamefont {Hemmer}}, \ and\ \bibinfo {author}
  {\bibfnamefont {M.}~\bibnamefont {Lukin}},\ }\href
  {http://dx.doi.org/10.1126/science.1131871} {\bibfield  {journal} {\bibinfo
  {journal} {Science}\ }\textbf {\bibinfo {volume} {314}},\ \bibinfo {pages}
  {281} (\bibinfo {year} {2006})}\BibitemShut {NoStop}%
\bibitem [{\citenamefont {Epstein}\ \emph {et~al.}(2005)\citenamefont
  {Epstein}, \citenamefont {Mendoza}, \citenamefont {Kato},\ and\ \citenamefont
  {Awschalom}}]{Epstein2005}%
  \BibitemOpen
  \bibfield  {author} {\bibinfo {author} {\bibfnamefont {R.~J.}\ \bibnamefont
  {Epstein}}, \bibinfo {author} {\bibfnamefont {F.~M.}\ \bibnamefont
  {Mendoza}}, \bibinfo {author} {\bibfnamefont {Y.~K.}\ \bibnamefont {Kato}}, \
  and\ \bibinfo {author} {\bibfnamefont {D.~D.}\ \bibnamefont {Awschalom}},\
  }\href {http://dx.doi.org/10.1038/nphys141} {\bibfield  {journal} {\bibinfo
  {journal} {Nat Phys}\ }\textbf {\bibinfo {volume} {1}},\ \bibinfo {pages}
  {94} (\bibinfo {year} {2005})}\BibitemShut {NoStop}%
\bibitem [{\citenamefont {Gaebel}\ \emph {et~al.}(2006)\citenamefont {Gaebel},
  \citenamefont {Domhan}, \citenamefont {Popa}, \citenamefont {Wittmann},
  \citenamefont {Neumann}, \citenamefont {Jelezko}, \citenamefont {Rabeau},
  \citenamefont {Stavrias}, \citenamefont {Greentree}, \citenamefont {Prawer}
  \emph {et~al.}}]{Gaebel2006}%
  \BibitemOpen
  \bibfield  {author} {\bibinfo {author} {\bibfnamefont {T.}~\bibnamefont
  {Gaebel}}, \bibinfo {author} {\bibfnamefont {M.}~\bibnamefont {Domhan}},
  \bibinfo {author} {\bibfnamefont {I.}~\bibnamefont {Popa}}, \bibinfo {author}
  {\bibfnamefont {C.}~\bibnamefont {Wittmann}}, \bibinfo {author}
  {\bibfnamefont {P.}~\bibnamefont {Neumann}}, \bibinfo {author} {\bibfnamefont
  {F.}~\bibnamefont {Jelezko}}, \bibinfo {author} {\bibfnamefont
  {J.}~\bibnamefont {Rabeau}}, \bibinfo {author} {\bibfnamefont
  {N.}~\bibnamefont {Stavrias}}, \bibinfo {author} {\bibfnamefont
  {A.}~\bibnamefont {Greentree}}, \bibinfo {author} {\bibfnamefont
  {S.}~\bibnamefont {Prawer}},  \emph {et~al.},\ }\href
  {http://dx.doi.org/10.1038/nphys318} {\bibfield  {journal} {\bibinfo
  {journal} {Nature Physics}\ }\textbf {\bibinfo {volume} {2}},\ \bibinfo
  {pages} {408} (\bibinfo {year} {2006})}\BibitemShut {NoStop}%
\bibitem [{\citenamefont {Fuchs}\ \emph {et~al.}(2008)\citenamefont {Fuchs},
  \citenamefont {Dobrovitski}, \citenamefont {Hanson}, \citenamefont {Batra},
  \citenamefont {Weis}, \citenamefont {Schenkel},\ and\ \citenamefont
  {Awschalom}}]{Fuchs2008}%
  \BibitemOpen
  \bibfield  {author} {\bibinfo {author} {\bibfnamefont {G.~D.}\ \bibnamefont
  {Fuchs}}, \bibinfo {author} {\bibfnamefont {V.~V.}\ \bibnamefont
  {Dobrovitski}}, \bibinfo {author} {\bibfnamefont {R.}~\bibnamefont {Hanson}},
  \bibinfo {author} {\bibfnamefont {A.}~\bibnamefont {Batra}}, \bibinfo
  {author} {\bibfnamefont {C.~D.}\ \bibnamefont {Weis}}, \bibinfo {author}
  {\bibfnamefont {T.}~\bibnamefont {Schenkel}}, \ and\ \bibinfo {author}
  {\bibfnamefont {D.~D.}\ \bibnamefont {Awschalom}},\ }\href
  {http://dx.doi.org/10.1103/PhysRevLett.101.117601} {\bibfield  {journal}
  {\bibinfo  {journal} {Phys. Rev. Lett.}\ }\textbf {\bibinfo {volume} {101}},\
  \bibinfo {pages} {117601} (\bibinfo {year} {2008})}\BibitemShut {NoStop}%
\bibitem [{\citenamefont {Ziegler}(2010)}]{Ziegler2010}%
  \BibitemOpen
  \bibfield  {author} {\bibinfo {author} {\bibfnamefont {J.~F.}\ \bibnamefont
  {Ziegler}},\ }\href {http://www.srim.org} {\emph {\bibinfo {title} {The
  Stopping and Range of Ions in Matter}}}\ (\bibinfo {year} {2010})\BibitemShut
  {NoStop}%
\bibitem [{\citenamefont {Toyli}\ \emph {et~al.}(2010)\citenamefont {Toyli},
  \citenamefont {Weis}, \citenamefont {Fuchs}, \citenamefont {Schenkel},\ and\
  \citenamefont {Awschalom}}]{Toyli2010}%
  \BibitemOpen
  \bibfield  {author} {\bibinfo {author} {\bibfnamefont {D.~M.}\ \bibnamefont
  {Toyli}}, \bibinfo {author} {\bibfnamefont {C.~D.}\ \bibnamefont {Weis}},
  \bibinfo {author} {\bibfnamefont {G.~D.}\ \bibnamefont {Fuchs}}, \bibinfo
  {author} {\bibfnamefont {T.}~\bibnamefont {Schenkel}}, \ and\ \bibinfo
  {author} {\bibfnamefont {D.~D.}\ \bibnamefont {Awschalom}},\ }\href
  {hppt://dx.doi.org/10.1021/nl102066q} {\bibfield  {journal} {\bibinfo
  {journal} {Nano Letters}\ }\textbf {\bibinfo {volume} {10}},\ \bibinfo
  {pages} {3168} (\bibinfo {year} {2010})}\BibitemShut {NoStop}%
\bibitem [{\citenamefont {Karrai}\ and\ \citenamefont
  {Tiemann}(2000)}]{Karrai2000}%
  \BibitemOpen
  \bibfield  {author} {\bibinfo {author} {\bibfnamefont {K.}~\bibnamefont
  {Karrai}}\ and\ \bibinfo {author} {\bibfnamefont {I.}~\bibnamefont
  {Tiemann}},\ }\href {hppt://dx.doi.org/10.1103/PhysRevB.62.13174} {\bibfield
  {journal} {\bibinfo  {journal} {Phys. Rev. B}\ }\textbf {\bibinfo {volume}
  {62}},\ \bibinfo {pages} {13174} (\bibinfo {year} {2000})}\BibitemShut
  {NoStop}%
\bibitem [{\citenamefont {Drexhage}(1974)}]{Drexhage1974}%
  \BibitemOpen
  \bibfield  {author} {\bibinfo {author} {\bibfnamefont {K.~H.}\ \bibnamefont
  {Drexhage}}\ }(\bibinfo  {publisher} {Elsevier},\ \bibinfo {year} {1974})\
  pp.\ \bibinfo {pages} {163 -- 192, 192a, 193--232}\BibitemShut {NoStop}%
\end{thebibliography}%

\section*{Methods}

\subsection{Diamond tip fabrication.}

Devices were fabricated from a sample of high purity, single crystalline diamond (Element Six, electronic grade) of $50~\mu$m thickness. We implanted the sample with atomic nitrogen at an energy and density of $6~$keV and $3\cdot10^{11}~$cm$^{-2}$, respectively. Subsequent annealing at $800^\circ$C for two hours yielded a shallow layer of NV centers of density ($\approx 10~$NVs/$\mu$m$^2$), and depth $\approx10~$nm. We then etched the sample from the back-side to a thickness $\approx3~\mu$m with reactive ion etching (RIE, Unaxis shuttleline), using a combined ArCl$_2$\,\cite{Lee2008} and O$_2$\,\cite{Hausmann2010} process. On the thin diamond membrane, we fabricated an array of diamond nanopillars on the top-side by using electron-beam lithography and RIE as described in\,\cite{Hausmann2010}. Next, we performed a second lithography step on the back-side of the diamond slab, which defined platforms to hold the diamond nanopillars. A final RIE process transferred the resist-pattern to the sample, and fully cut through the diamond membrane to yield in the structure shown in Fig.\,\ref{figureSchematics}d.


To mount a pre-selected diamond platform on an AFM tip, we employed a focussed ion beam (FIB) system (Zeiss NVision 40) which was equipped with a nanomanipulator (Omniprobe AutoProbe $300$) and ion-assisted metal deposition. We employed tungsten deposition to fuse a diamond platform to a quartz AFM tip and then used FIB cutting to release the diamond platform from the bulk. With a properly aligned FIB, this process does not contaminate the scanning diamond nanopillar, and yields a NV/AFM probe as shown in Fig.\,\ref{figureSchematics}b.

\subsection{Combined confocal and atomic-force microscope.}

We employed a homebuilt microscope combining optical (confocal) imaging and AFM. The optical microscope is based on a long working-distance microscope objective (Mitutoyo ULWD HR NIR 100x, 0.7NA). The AFM was tuning-fork based, controlled using commercial electronics (Attocube ASC500) and mounted using a home-built AFM-head. Both the sample and the AFM-head were fixed on 3-axis coarse and fine positioning units (Attocube ANPxyz101 and ANSxyz100, respectively) to allow positioning of the diamond tip with respect to the fixed optical axis and subsequent scanning of the sample with respect to the diamond probe.

Optical excitation of the NV center was performed by a diode-pumped solid-state laser (LaserGlow LRS-0532-PFM-00100-01) at a wavelength of $532~$nm. Pulsed excitation for coherent NV spin manipulation used a double-pass acousto-optical modulator AOM setup (Isomet, AOM 1250C-848). ESR  was driven with a microwave generator (Rhode Schwartz, SMB100A) and amplifier (MiniCircuits, ZHL-42W). Both the AOM and microwave source were timed using a computer-controlled trigger-card (Spincore, PulseBlasterESR-PRO-400).

\subsection{Fit to spin-echo data.}

To obtain the NV $T_2$-time form the spin-echo measurement presented in Fig.\,\ref{figureBasicProperties}d, we fitted the data to a sum of gaussian peaks, modulated by a decay envelope $\propto{\rm exp}[-(\tau/T_2)^n]$, i.e., we employed the fit-function\,\cite{Childress2006}

\begin{equation}
{\rm exp}[-(\tau/T_2)^n]\sum\limits_{j} {\rm exp}[-((\tau-j\tau_{\rm rev})/T_{\rm dec})^2].
\end{equation}

Taking $T_2,n,\tau_{\rm rev}$ and $T_{\rm dec}$ as free fitting parameters, we found $T_2=33.4~\mu$s, $n=1.4$, $\tau_{\rm rev}=16~\mu$s and $T_{\rm dec}=4.9~\mu$s for the data shown in Fig.\,\ref{figureBasicProperties}d.

The following supplementary material is divided into five sections. Each section provides background information related to specific topics of the main text. The sections are not built upon each other and can be read independently. \
Section\,\ref{SectSim} provides details for the model-calculation used to simulate the NV magnetic image in Fig.3d.
In section\,\ref{SectMagQuench}, we discuss limitations to NV magnetic imaging if the NV sensor is in close proximity to a strongly magnetized sample.
Experimental limitations to the achievable NV-to-sample distance are discussed in section\,\ref{SectLimitDist}.
The fabrication of the sharp metallic tips employed in FQI (Fig.4a) is detailed in section\,\ref{SectFabTips}.
Finally, section\,\ref{SectFQI} contains a description and simple model for the topographic features observed in the FQI image in Fig.4a.

\newpage

\section*{Supplementary Information}

The following supplementary material is divided into five sections. Each section provides background information related to specific topics of the main text. The sections are not built upon each other and can be read independently. \
Section\,\ref{SectSim} provides details for the model-calculation used to simulate the NV magnetic image in Fig.3d.
In section\,\ref{SectMagQuench}, we discuss limitations to NV magnetic imaging if the NV sensor is in close proximity to a strongly magnetized sample.
Experimental limitations to the achievable NV-to-sample distance are discussed in section\,\ref{SectLimitDist}.
The fabrication of the sharp metallic tips employed in FQI (Fig.4a) is detailed in section\,\ref{SectFabTips}.
Finally, section\,\ref{SectFQI} contains a description and simple model for the topographic features observed in the FQI image in Fig.4a.

\subsection{S1. Simulation of magnetic images}
\label{SectSim}

\begin{figure*}
\includegraphics{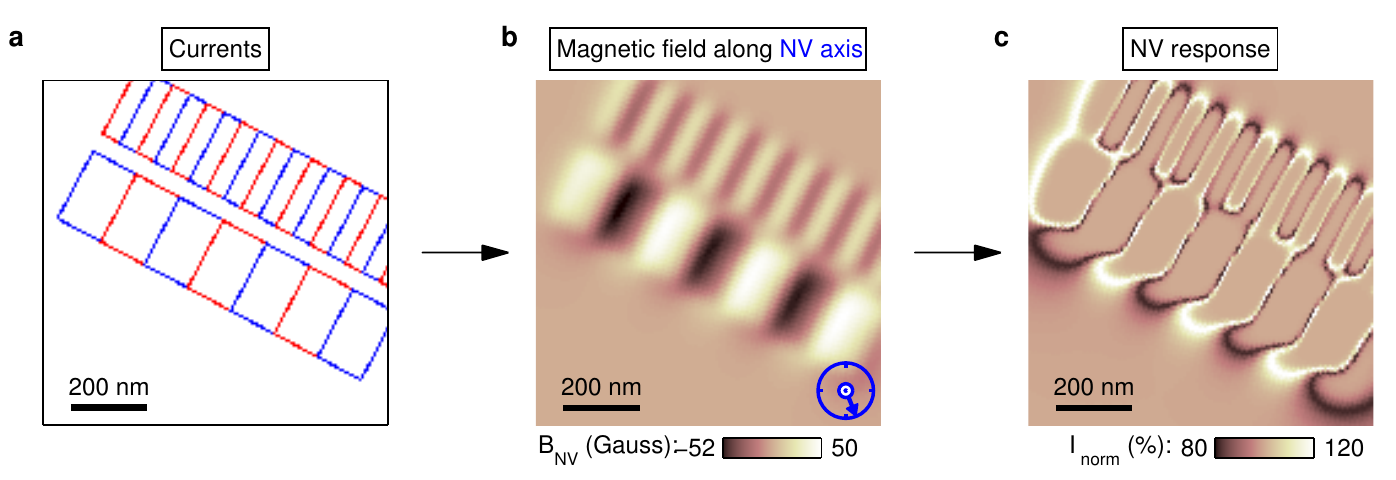}
\caption{\label{figureMagSim} \textbf{Simulation of NV response to bits of a magnetic memory}. (a) Current distribution used to simulate the magnetic bits imaged in this work. Red (blue) loops indicate currents of $1~$mA in the (counter-)clockwise direction. (b) Magnetic field generated by the current-distribution in (a), projected on the NV axis at a height of $50~$nm above the current loops. The NV axis was tilted by $37^\circ$ out of the scan plane ($[0 \overline{1} 1]$ crystalline direction) with an in-plane component as illustrated by the blue arrow. (c) NV magnetometry response obtained from the magnetic field distribution in (b), assuming a Lorentzian NV-ESR response and RF detunings as in the original experiment (see text).}
\end{figure*}

In order to reproduce the magnetic images obtained with the scanning NV sensor, we performed a model-calculation of the local magnetic fields in proximity to the hard-disc sample we imaged in our experiment. The magnetic domains were approximated by an array of current-loops in the sample-plane as illustrated in Fig.\,\ref{figureMagSim}a. We chose the sizes of the loops to match the nominal size of the magnetic bits on the sample (bit-with $200~$nm and bit-length $125~$nm and $50~$nm for the tracks in the figure) and set the current to $1~$mA (corresponding to a density of $\approx1~$ Bohr magneton per $(0.1~{\rm nm})^2$), which we found to yield the best qualitative match to the magnetic field strengths observed in the experiment. We then applied Biot-Savart's law to this current-distribution to obtain the magnetic field distribution in the half-plane above the sample.

Fig.\,\ref{figureMagSim}b shows the resulting magnetic field projection onto the NV center at a scan height of $50~$nm above the current loops. The NV direction was experimentally determined to be along the ($[0 \overline{1} 1]$) crystalline direction of the diamond nanopillar (in a coordinate-system where $x-$, $y-$ and $z-$ correspond to the horizontal-, vertical and out-of plane directions in Fig.\,\ref{figureMagSim}b), by monitoring the NV-ESR response to an externally applied magnetic field (using 3-axis Helmholtz-coils). We then allowed for slight variations of the NV orientation due to alignment errors between the diamond crystallographic axes and the scan directions to find the NV orientation that reproduced our experimental data best. With this procedure, we found an NV orientation $(\sqrt{2}{\rm sin}(\phi),\sqrt{2}{\rm cos}(\phi),1)/\sqrt{5}$, with $\phi=\pi 162/180$.

Finally, we used this magnetic-field distribution to calculate the response of the NV center to a magnetometry scan as described in the main text. For this, we assumed a Lorentzian ESR response with a full-width at half maximum of $9.7~$MHz, a visibility of $20~$\% and two external RF sources with detunings $\pm10~$MHz from the bare ESR frequency, all in accordance with our original experimental parameters.

\subsection{S2. NV magnetometry in close proximity to a strongly magnetized sample}
\label{SectMagQuench}

\begin{figure*}
\includegraphics{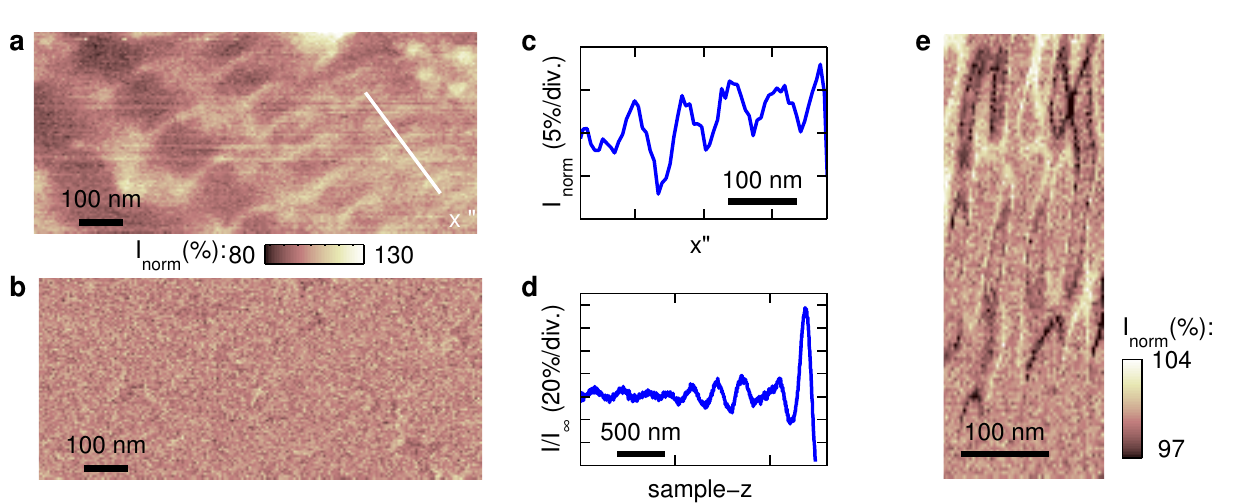}
\caption{\label{figureHDQuench} \textbf{Quenching of NV fluorescence and ESR contrast in hard-disc imaging}. (a) Total NV fluorescence $I_{\rm norm}$ as a function of sample position for an NV in close proximity to the hard-disc sample. $I$ was normalized to the average fluorescence intensity of $I_0\approx15000~$cps in the scan. Dark regions in the scan correspond to individual magnetic domains and are caused by strong magnetic fields transverse to the NV axis which occur in close proximity to the domains.  (b) NV magnetic image recorded simultaneously with (a). Data acquisition and intrgration time per pixel was analogous to the magnetic imaging described in the main text. However here, due to strong transverse magnetic fields, NV ESR contrast almost completely disappeared and prevented NV magnetic imaging using optically detected ESR. The color-bar applies to (a) and (b). (c) Line-cut along the white line in (a), averaged over $7$ adjacent pixels. $I_{\rm norm}$ shows a periodicity of $\approx64~$nm, indicating a bit-width of $32~$nm. (d) Fluorescence approach curve on the magnetic memory medium. NV fluorescence $I$ was normalized to the fluorescence rate $I_\infty=27'000~$cps when the NV center was far from the sample. In contact with the magnetic sample, NV fluorescence was reduced by almost a factor of two compared to the NV counts far from the sample. (e) Magnetic imaging with the same NV sensor: Even in close contact to the sample, NV magnetic imaging using ESR is still possible, albeit with a strongly reduced ESR contrast and signal to noise ratio compared to the data shown in the main text. Data in (e) was acquired over $180~$minutes, the smallest resolvable magnetic domains (top third of image) have a width of $\approx~20~$nm.}
\end{figure*}

The presence of a strong magnetic field $B_\bot$, transverse to the NV axis leads to a reduction of contrast in optically detected ESR and moreover reduces the overall fluorescence intensity of the NV center\,\cite{Epstein2005}. These effects result from a mixing of the NV spin-levels in the optical ground and excited states of the NV center in the presence of $B_\bot$. Such mixing on one hand allows for spin non-conserving optical transitions and on the other hand suppresses the spin-dependance in shelving from the NV excited state (triplet) to the metastable NV singlet state. Both, spin-conservation under optical excitation and spin-dependant shelving are responsible for the non-zero contrast in optically detected ESR of NV centers\,\cite{Gaebel2006} and consequently, their suppression with transverse magnetic fields explains the disappearance of NV magnetometry features when closely approaching a strongly magnetized sample.

Fig.\,\ref{figureHDQuench}a shows the raw NV fluorescence counts observed when scanning an NV in a diamond nanopillar in close proximity (estimated $10-20~$nm distance between NV and sample surface) to the sample. Dark features appear when the NV is scanned over magnetic bits that enhance $B_\bot$, while the inverse happens when $B_\bot$ is reduced (or the longitudinal field $B_{\rm NV}$ enhanced) by local fields. This mode of bit-imaging allows for spatial resolutions $\approx20-30~$nm (Fig.\,\ref{figureHDQuench}c). At the same time, a magnetic image recorded with the technique described in the main text shows no appreciable imaging contrast (Fig.\,\ref{figureHDQuench}b). Only exceedingly long integration times on the order of hours allowed us to reveal weak magnetic features with dimensions on the order of $20~$nm (Fig.\,\ref{figureHDQuench}d).

The rates of the two effects which lead to a disappearance of ESR contrast, i.e. spin-flip optical transitions and shelving of $m_s=0$ electronic states into the metastable singlet, scale approximately as $\left(\frac{B_\bot}{D_{\rm GS}-D_{\rm ES}}\right)^2$ and $\left(\frac{B_\bot}{D_{\rm ES}}\right)^2$, respectively, with $D_{\rm GS (ES)}$ the ground- (excited-) state zero-field spin-splitting of $2.87~$GHz and $1.425~$GHz\,\cite{Fuchs2008}, respectively. Given that $D_{\rm GS}\approx 2 D_{\rm ES}$, the scaling of the two mechanisms with $B_\bot$ will be very similar. The characteristic scale of $D_{\rm ES}$ ($D_{\rm GS}/2$) for the disappearance of ESR contrast thus allows us to estimate $B_\bot$ close to the sample to be $B_\bot\approx D_{\rm ES}/\gamma_{\rm NV}\approx 514~$Gauss. We note however that this simple argument likely gives and over-estimation of $B_\bot$ as smaller values can already significantly effect ESR contrast and NV fluorescence intensity due to the complex dynamics of NV spin pumping. Indeed, strong reductions of NV fluorescence rates for $B_\bot$ less than $100~$G have been observed in the past\,\cite{Epstein2005}. Transverse magnetic fields on this order were consistent with the largest \emph{on-axis} magnetic fields observed on our experiments as well as with the calculations of magnetic field profiles presented in Sect.\,\ref{SectSim} (for the parameters used in Fig.\,\ref{figureMagSim}, we obtain maximal values of $B_\bot\approx200~$Gauss for an NV-to-sample distance of $20~$nm).

\subsection{S3. Limitations to NV-sample distance}
\label{SectLimitDist}

\begin{figure*}
\includegraphics{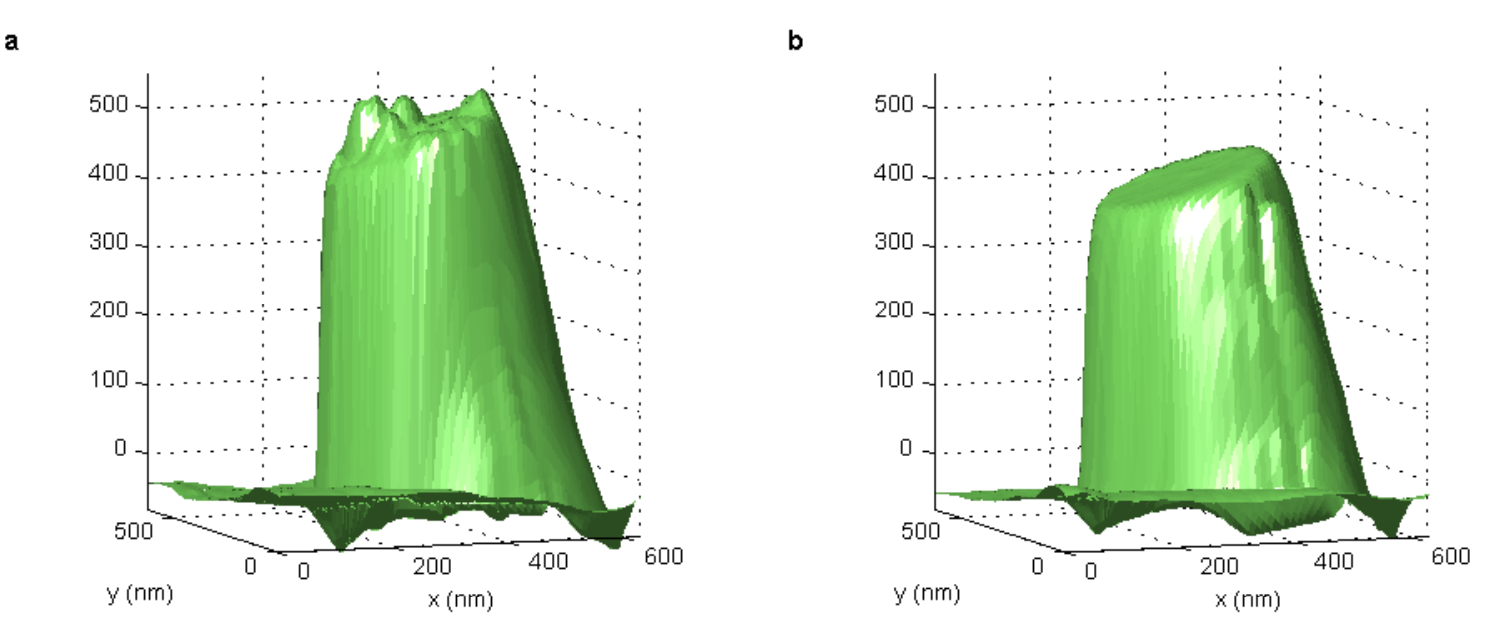}
\caption{\label{figureTipGunk} \textbf{Contamination of diamond tips}. (a) AFM image of the end of a scanning diamond nanopillar after contamination during scanning. The image was acquired by scanning the diamond nanopillar over a sharp diamond tip as shown in Fig.\,\ref{figureSharptip}. (b) AFM Image of the same nanopillar as in (a) after cleaning of the pillar's end-face by repeated ``scratching'' over the sharp diamond tip.}
\end{figure*}

As mentioned in the main text, NV-sample distance is an essential parameter for the performance of our microscope as it determines the overall resolving power with which weak magnetic targets can be imaged. We identified three critical parameters that can affect NV-sample distance:

\begin{itemize}

\item Implantation-depth of NV centers in the diamond nanopillars

The depth of the NV centers created using ion implantation can be controlled by the energy of the ions used for NV creation. However, the stopping of ions in matter is a random process\,\cite{Ziegler2010} and the depth of the created NV centers therefore not perfectly well-defined. This straggle in ion implantation poses an intrinsic uncertainty to the distance between the scanning NV and the end of the diamond nanopillar. For implantation energies of $6~$keV (with nominal implantation-depths of $10~$nm) as used in this work, NV straggle has recently been shown to be as large as $10-20~$nm\,\cite{Toyli2010,Grinolds2011}. We note that since straggle in NV implantation is hard to circumvent it is essential for the future to develop techniques to precisely pre-determine the depth of a given sensing NV in a diamond nanopillar. This could be performed using recently developed nanoscale imaging methods for NV centers\,\cite{Grinolds2011}, or by scanning the NV sensor over a well-defined magnetic field source.

\item Contamination of scanning diamond nanopillars

During scanning-operation, the scanning diamond nanopillar can gather contamination from the sample or environment. An example for such a contaminated diamond-tip is shown in the AFM image shown in Fig.\,\ref{figureTipGunk}a (which was acquired with the scanning protocol employed in Fig.4, using the a sharp diamond tip as shown in Fig.\,\ref{figureSharptip}). Such contamination can artificially increase the distance of the scanning NV center to the sample by several $10~$'s of nm (see Fig.\,\ref{figureTipGunk}a). To undo contamination of the diamond-tip after excessive scanning over dirty samples, we developed a ``tip-cleaning technique'' that allowed us to revert a contaminated tip to its initial, clean state (as illustrated by the transition from Fig.\,\ref{figureTipGunk}a to b). Tip cleaning is performed by repeated scanning of the diamond nanopillar over a sharp diamond tip (Fig.\,\ref{figureTipGunk}a) in the absence of AFM feedback. Such feedback-free scanning can partly remove contamination from the diamond pillar, which after repeated operation leads to a clean device as the one shown in Fig.\,\ref{figureTipGunk}b.

We note that with proper sample-cleaning, control over environmental conditions and occasional ``tip-cleaning'' runs, adverse effects of tip-contamination can be essentially eliminated. This, together with the excellent photo-stability of NV centers, then allows for long-term operation of the scanning NV sensor.

\item AFM control

Proper AFM control is necessary to assure close proximity of the NV center to the sample surface. It has been shown in the past that bad mounting or improper AFM feedback control can lead to AFM tip-sample distances in excess of $20~$nm\,\cite{Karrai2000}. Careful mounting of AFM tips and proper setup and tuning of AFM feedback (here provided by an Attocube ASC500 controller) was therefore essential to observe, for instance, the FQI features discussed in Fig.4 of the main text.

\end{itemize}

\subsection{S4. Fabrication of sharp diamond tips}
\label{SectFabTips}

\begin{figure}
\includegraphics{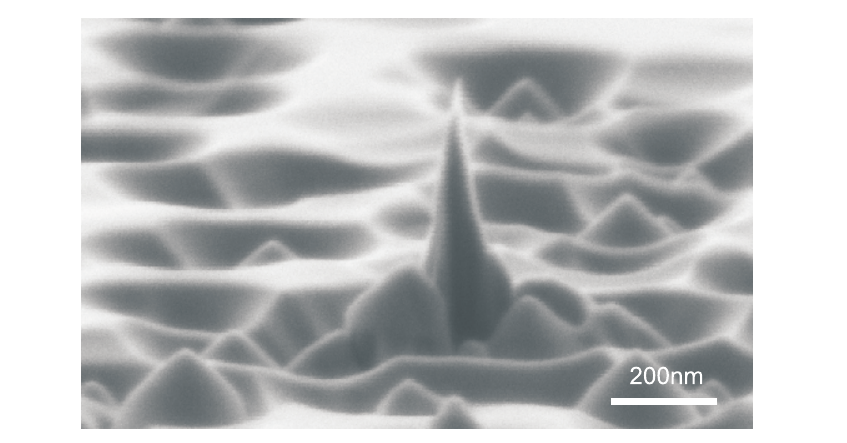}
\caption{\label{figureSharptip} \textbf{Sharp diamond tip for FQI}. Image of a sharp diamond tip similar to the one used for the experiments presented in Fig.4a of the main text. Typical tip-radii are on the order of 10nm.}
\end{figure}

For the experiment presented in Fig.4a of the main text, we fabricated sharp diamond tips which were metal coated for FQI. Diamond tip fabrication was based on the nanofabrication techniques\,\cite{Hausmann2010} that we already employed for the production of the scanning diamond nanopillars presented in Fig.1. A type Ib diamond (Element six) was patterned with circular etch-masks (flowable oxide, FOx XR, Dow Corning) of $100~$nm diameter. Here, in order to obtain sharp diamond tips instead of cylindrical diamond nanopillars, we modified the RIE etching recipe we had previously used: While we kept the (oxygen) etching chemistry identical to pillar fabrication, we significantly increased the etching time, such as to completely erode the etch mask on the diamond substrate. As a result, the etched diamond structures acquired the form of sharp tips as shown in the representative SEM image in Fig.\,\ref{figureSharptip}. Typical tip-radii were in the range of $10~$nm and tip lengths were on the order of $200~$nm.

For FQI, we then coated the sharp diamond tips with a thin metallic layer using thermal metal evaporation. To avoid oxidation of the metal, we chose gold as the quenching metal and used a chrome adhesion layer between the gold and the diamond. For the tips employed in this work, we deposited $5~$nm of gold and $5~$nm of chrome.

\subsection{S5. Explanation of FQI features}
\label{SectFQI}

\begin{figure*}
\includegraphics{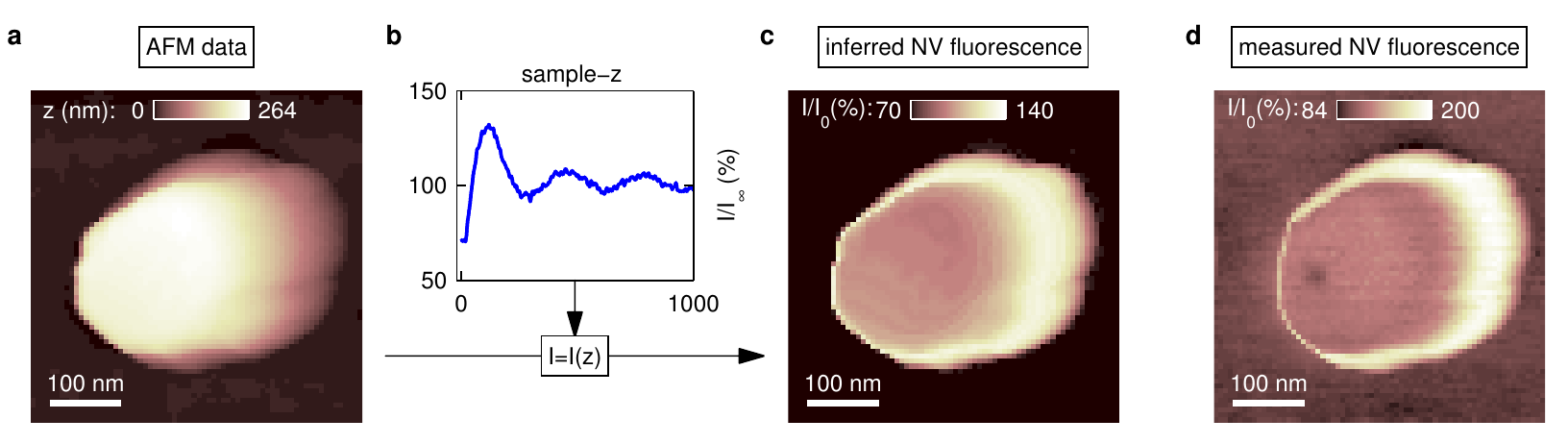}
\caption{\label{figureFluoQuench} \textbf{Explanation of topographic features in FQI}. (a) AFM topography recorded during FQI imaging (same data as shown in Fig.4c). (b) Fluorescence ``approach-curve'' of the FQI sample to the NV center. (c) FQI image reconstructed form the datasets in (a) and (b): Looking up the NV fluorescence intensity in (b) for every tip-sample displacement measured in (a) yields the reconstructed topographic features in FQI shown in the panel. (d) Original FQI data (same data as Fig.4a). The features common to (c) and (d) are attributed to effects of sample topography. The additional, dark feature in the center of (d) (red square in Fig.4a) has no correspondence in topography and stems from direct FQI of the NV center on the sharp metallic tip.}
\end{figure*}

The features observed in Fig.4a of the main text were governed by direct fluorescence quenching through metallic objects (as highlighted by the red square in the figure) and by a confluence of the distance-dependance of the NV fluorescence with topographic features on the sample (bright, ring-shaped feature in the figure). When approaching the NV to the (metallic) sample, the total NV fluorescence collected in the far-field through the pillar changed as shown in the measurement in Fig.4e (blue curve) and the corresponding data for the FQI sample shown in Fig.\,\ref{figureFluoQuench}b. This well-known\,\cite{Drexhage1974} variation of NV fluorescence is a result of the variable electromagnetic density of states in the vicinity of a dielectric interface which influences the NV radiative lifetime as well as the total effective laser excitation intensity impinging on the NV center. During our scanning experiments, the topography causes the mean distance between the scanning NV center in the nanopillar and the metallic substrate to vary, which in turn causes variations in the collected NV fluorescence rate. Assuming to first order that the metallic tip does not itself affect NV fluorescence (so long as it is not placed in direct contact to the NV center as in the ``red-square region''), one can understand most features observed in Fig.4c as a pure effect of topography. Based on this principle, in Fig.\,\ref{figureFluoQuench} we reconstruct the FQI image from a measurement of sample topography (a) and an independently acquired fluorescence ``approach-curve'' (b). The reconstructed FQI image (Fig.\,\ref{figureFluoQuench}c) was obtained by taking the value of the AFM z-displacement for each point in the scan and looking up the corresponding fluorescence-rate obtained in the approach-curve. The resulting image shows striking similarity with the actually measured FQI image (Fig.\,\ref{figureFluoQuench}e; same data as Fig.4a) and confirms the validity of our explanation.

\end{document}